\documentclass[letter,11pt]{article}
\pdfoutput=1 % if your are submitting a pdflatex (i.e. if you have
\usepackage{jcappub} % for details on the use of the package, please
\usepackage[T1]{fontenc} % if needed

\usepackage{amssymb}
\usepackage{graphicx, epsfig, bm, amsmath}
\usepackage{hyperref}
\usepackage{subcaption}
\usepackage{soul}

\usepackage{color}
\usepackage{ifthen}
\def\nn{\nonumber}
\def\({\left(}
\def\){\right)}
\def\[{\left[}
\def\]{\right]}
\def\tr{{\rm Tr}}
\def\gfc{\xi}
\newcommand{\beq}{\begin{equation}}
\newcommand{\beqn}{\begin{eqnarray}}
\newcommand{\eeq}{\end{equation}}
\newcommand{\eeqn}{\end{eqnarray}}

\bibstyle{JHEP}

\title{Magnetogenesis from axion inflation}

\author[a]{Peter Adshead,}
\author[b,c]{John T. Giblin, Jr.}
\author[a]{Timothy R. Scully,}
\author[a]{Evangelos I. Sfakianakis}

\affiliation[a]{Department of Physics, University of Illinois at Urbana-Champaign, Urbana, Illinois 61801, U.S.A.}
\affiliation[b]{Department of Physics, Kenyon College, Gambier, Ohio 43022, U.S.A.}
\affiliation[c]{Department of Physics, Case Western Reserve University, Cleveland, Ohio 44106, U.S.A.}

\emailAdd{adshead@illinois.edu}
\emailAdd{giblinj@kenyon.edu}
\emailAdd{tscully2@illinois.edu}
\emailAdd{esfaki@illinois.edu}

\abstract{
In this work we compute the production of magnetic fields in models of axion inflation coupled to the hypercharge sector of the Standard Model through a Chern-Simons interaction term. We make the simplest choice of a quadratic inflationary potential and use lattice simulations to calculate the magnetic field strength, helicity and correlation length at the end of inflation. For small values of the axion-gauge field coupling strength the results agree with no-backreaction calculations and estimates found in the literature. For larger couplings the helicity of the magnetic field differs from the no-backreaction estimate and depends strongly on the comoving wavenumber. We estimate the post-inflationary evolution of the magnetic field based on known results for the evolution of helical and non-helical magnetic fields. The magnetic fields produced by axion inflation with large couplings to $U(1)_Y$ can reach $B_{\rm eff} \gtrsim 10^{-16}\, {\rm G}$, exhibiting a field strength $B_{\rm phys} \approx 10^{-13}\, {\rm G}$ and a correlation length $\lambda_{\rm phys}\approx10\, {\rm pc}$. This result is insensitive to the exact value of the coupling, as long as the coupling is large enough to allow for instantaneous preheating. 
Depending on the assumptions for the physical processes that determine blazar properties, these fields can be found consistent with blazar observations based on the value of $B_{\rm eff}$. Finally, the intensity of the magnetic field for large coupling can be enough to satisfy the requirements for a recently proposed baryogenesis mechanism, which utilizes the chiral anomaly of the Standard Model.
\\
}

\begin{document}
\maketitle
\flushbottom

%%%%%%%%%%%%%%%%%%%%%%%%%%%%%%%%%%%%%%%%%%%%%%%%%%%
%%%%%%%%%%%%%%%%%%%%%%%%%%%%%%%%%%%%%%%%%%%%%%%%%%%
%%%%%%%%%%%%%%%%%%%%%%%%%%%%%%%%%%%%%%%%%%%%%%%%%%%
%%%%%%%%%%%%%%%%%%%%%%%%%%%%%%%%%%%%%%%%%%%%%%%%%%%

\section{Introduction}
\label{sec:intro}

Magnetic fields appear to be ubiquitous in our Universe. They have been measured at length scales that range from those within our solar system and other single star systems to galaxies and galaxy clusters \cite{Widrow:2002ud,Bernet:2008qp,Beck:2008ty,Kronberg:2007dy}. Furthermore, observations of distant blazars point towards the existence of magnetic fields in intergalactic voids \cite{Neronov:1900zz, Tavecchio:2010mk, Dolag:2010ni, Essey:2010nd, Taylor:2011bn, Takahashi:2013lba, Finke:2013tyq, Finke:2015ona}. The strength of the observed field are weakly scale dependent. Galactic magnetic fields have strengths of up to tens of $\mu$Gauss, while those of galaxy clusters are of $\mu$Gauss strength. The origin of these magnetic fields is a long standing problem in cosmology. The standard hypothesis to explain galactic fields is the action of the dynamo mechanism. However, the dynamo can only amplify existing magnetic fields \cite{Brandenburg:2004jv}, not create them {\sl ex nihilo}. The amplitude required for these seed fields depends on the scale as well as the details of the amplifying dynamo.  The magnitude of these magnetic fields is somewhat degenerate with the corresponding correlation length. The quantity that can be constrained by observations is $B_{\rm eff} \ge 10^{-15}$ G,  which is equal to the magnitude of the magnetic field when the correlation length is larger than $1$ Mpc. For a smaller correlation length $\lambda <1$ Mpc, the magnetic field is enhanced by $B = B_{\rm eff} \sqrt{1\,{\rm Mpc} / \lambda}$. This bound can be relaxed to $B_{\rm eff} \ge 10^{-17}$ G, depending on the assumptions made that suppress the cascade emission \cite{Taylor:2011bn}. 

Recent studies \cite{Tashiro:2013ita,Chen:2014qva} have used the diffuse part of the GeV photon spectrum measured by the Fermi Satellite in order to better probe the spectrum of intergalactic magnetic fields. The analysis uses parity-odd correctors of the diffuse gamma ray signal to extract information about the helicity of cosmological magnetic fields along with their strength {at cosmological distances}. While the reconstruction of the magnetic helicity spectrum relies on assumptions about the physical processes involved, such as the nature of the charged particles involved in the inverse Compton scattering of the CMB photons \cite{Chen:2014qva}, 
the amplitdude of the magnetic field at a scale of $ 10\, {\rm Mpc}$ can be inferred from the Fermi data to be $B \sim 5.5 \times 10^{-14}\, \rm{G}$.

Many attempts have been made to propose a primordial origin of large-scale magnetic fields (see for example \cite{Kandus:2010nw}). The main categories of models include magnetic field generation during (or immediately following) inflation \cite{Turner:1987bw, Ratra:1991bn, DiazGil:2007dy,Martin:2007ue,Demozzi:2009fu,Kanno:2009ei,Bamba:2006ga,Ng:2015ewp,Ferreira:2013sqa,Ferreira:2014hma,Campanelli:2013mea}, and early universe phase transitions (QCD and electroweak, for example) \cite{Vachaspati:1991nm, Sigl:1996dm, Dolgov:2001nv, Stevens:2007ep, Kahniashvili:2009qi, Henley:2010ba,Grasso:2000wj,Cheng:1994yr}. For recent reviews of primordial magnetic fields, and their cosmological evolution and detection, see ref.\ \cite{Durrer:2013pga,Widrow:2011hs} and references therein.

Inflationary magnetogenesis models also exhibit significant variation. Since Maxwell's action is conformally invariant, there can be no significant magnetic field production during inflation.   Without attempting to provide an exhaustive categorization of methods to break conformal invariance during inflation, we can point at two main families of models, both consisting of a standard slow-rolling inflaton and a $U(1)$ gauge field  that can be assumed to be, or easily translated to, the electromagnetic field. The coupling between the inflaton and gauge field distinguishes between the different models and can be taken to include a term of the form $V_{\rm interaction}=I(\phi) F_{\mu\nu} F^{\mu\nu} $ or $V_{\rm interaction}=I(\phi) F_{\mu\nu} \tilde F^{\mu\nu}$, where $I(\phi)$ is some function of the inflaton field $\phi$.  The former is usually referred to as the Ratra model \cite{Ratra:1991bn}. It is generally very hard to produce the required amplitude of seed magnetic fields in the context of the Ratra model without producing significant non-Gaussianities in the cosmic microwave background, or suffering from the strong coupling problem \cite{Barnaby:2012tk}. A method to evade these constraints by coupling the field tensor directly to the curvature was recently proposed in ref.\ \cite{Guo:2015awg}.

The addition of an axial coupling term $V_{\rm interaction}=I(\phi) F_{\mu\nu} \tilde F^{\mu\nu}$, coupling the gauge field to the axion results in the production of helical fields \cite{Campanelli:2008kh}. 
Helical magnetic fields produced in the early Universe have a better chance to exhibit a significant amplitude at large scales at the epoch of structure formation due to an effect known as {\it inverse cascade} \cite{Campanelli:2007tc,Banerjee:2004df}.
During the inverse cascade process, power is transferred from short- to long-wavelength modes, thereby protecting the magnetic field strength from decaying, and at the same time increasing the correlation length \cite{Son:1998my}.
For example, in the model proposed in ref.\ \cite{Caprini:2014mja}, which is a hybrid between the Ratra and axion model, helical magnetic fields are produced during inflation.
 In order to produce cosmologically relevant magnetic fields, inflation must occur well below the  inflationary scale that would give tensor modes near the current upper limit of the BICEP2 \& Planck analysis \cite{Array:2015xqh}. However, magnetic fields can source chiral gravitational waves, thereby leading to a large tensor-to-scalar ratio, $r$, and evading the Lyth bound.  The helical magnetic field generated during inflation by a pure Maxwell action and a coupling term $V_{\rm interaction}=I(\phi) F_{\mu\nu} \tilde F^{\mu\nu}$ was studied in \cite{Durrer:2010mq} and found to be insufficient for satisfying observational limits. 

The coupling of axions to gauge fields during inflation leads to rich phenomenology. Important aspects include the amplification of parity violating gauge fields during slow-roll inflation \cite{Carroll:1991zs, Garretson:1992vt, Ferreira:2014zia} and their influence on the inflationary dynamics \cite{Prokopec:2001nc, Anber:2009ua, Barnaby:2011vw, Barnaby:2011qe}, as well as the generation of
metric fluctuations by a rolling auxiliary pseudo-scalar during inflation \cite{Shiraishi:2013kxa, Cook:2013xea,Ferreira:2015omg,Peloso:2016gqs}. 
Recently models which employ larger couplings between the axion and gauge sectors have been proposed \cite{Anber:2009ua, Barnaby:2011qe}. Rescattering of the gauge fields off the axion condensate in these models can lead to observable effects during inflation, such as large non-Gaussianity \cite{Barnaby:2010vf}. The strength of the axion-gauge coupling is constrained by the requirement that effects such as non-Gaussianity of the density fluctuations, chiral gravitational waves, and the production of primordial black holes  do not exceed observational limits \cite{Barnaby:2010vf, Barnaby:2011vw, Barnaby:2011qe, Linde:2012bt, Bugaev:2013fya}, as well as by the limits of perturbation theory applied to the axion during inflation \cite{Ferreira:2015omg,Peloso:2016gqs}.

In ref.\ \cite{Adshead:2015pva} we used lattice simulations to study the transfer of energy from the axion-inflaton field to $U(1)$ gauge fields at the end of inflation. We found that, for reasonable ranges of the axion-gauge coupling, non-linear effects can be very important at the end of inflation. In particular, at the middle to upper range of the couplings allowed by black hole abundance,  a range where backreaction cannot be neglected and lattice simulations are essential, reheating is essentially instantaneous, proceeding via a phase of tachyonic resonance \cite{Dufaux:2006ee} and completing within a single oscillation of the axion. Despite the asymmetry in the equations of motion for the two polarizations of the gauge fields, rescattering of the gauge bosons off the axion condensate is efficient at generating the second polarization on sub-horizon scales. This can significantly reduce the helicity asymmetry of the resulting gauge field. On scales larger than the horizon at the end of inflation, an asymmetry between the gauge field polarizations remains. For these large couplings the Universe  is radiation dominated and  characterized by a high reheating temperature.   Even in the cases where preheating is not efficient, the axion-gauge field coupling provides a perturbative decay channel for the axion into gauge bosons. This guarantees that reheating will eventually complete through perturbative decays alone. Complete reheating is essential for the Universe to transition from inflation into the hot big bang.

Magnetogenesis via axial couplings to gauge fields has been previously studied in ref.\ \cite{Anber:2006xt} in the context of N-flation.  Previous studies of magnetogenesis from single-field axion inflation have focussed on the small-coupling regime where the backreaction of the gauge field on the inflaton can be neglected \cite{Garretson:1992vt}. This regime can be accurately modeled by solving the linear equations of motion for the fields and the background spacetime. The analysis of ref.\ \cite{Fujita:2015iga} took into account the backreaction of the gauge field to the inflaton in a Hartree-Fock type approximation. However, this analysis was used as a means to quantify the limit of the small backreaction regime, rather than an attempt to self-consistently simulate the system for large couplings (see also, ref.\ \cite{Cheng:2015oqa}).  Methods based on linear theory are questionable in the regions of parameter space we consider here, where the gauge fields significantly backreact on the inflaton.  

In this work we study the generation of helical magnetic fields during preheating and the viability of gauge fields produced during pseudo-scalar inflation to explain the origin of cosmologically relevant magnetic fields.  We perform lattice simulations to go well beyond the linear regime, and we calculate the strength and correlation length of the resulting primordial magnetic fields. By making some fairly generic assumptions about the reheating history and using standard results of the magneto-hydrodynamics (MHD) literature, we estimate the subsequent evolution of the inflationary magnetic fields. We demonstrate that magnetic fields can be generated with amplitudes sufficient to provide an explanation for blazar observations.

Our results can be briefly summarized as follows. 
In the low-coupling regime, the produced magnetic field is in agreement with simple no-backreaction calculation as well as results found in the literature. In the large-coupling regime, where the total energy density of the inflaton is transferred to the gauge field within a single inflaton background oscillation, the amplitude of the magnetic field reaches a maximum value of $B_{\rm phys}^2 \sim 10^9 m^4$, where $m$ is the inflaton mass. The magnetic field amplitude is largely insensitive to the exact value of the axion-gauge coupling, as long as it supports instantaneous preheating \cite{Adshead:2015pva} and is not too large, as to lead to prolonged inflation due to non-linear ``trapping'' of the axion.
Simple dimensional arguments show that scattering of the hypercharge gauge bosons into Standard model particles is efficient in the case of instantaneous preheating, filling the Universe with a charged plasma a few e-folds after the end of inflation. Inverse cascade processes of helical magnetic fields in a charged turbulent plasma can lead to a current magnetic field strength of $B_{\rm eff} \gtrsim 10^{-16 }\, G$, which is potentially consistent with astronomical observations of distant blazars. Furthermore, the intensity and correlation length of the resulting magnetic field can be relevant for a baryogenesis scenario proceeding through the chiral anomaly of the Standard Model \cite{Fujita:2016igl}.

The paper is structured as follows. In section \ref{background} we define the model and present the equations that govern the evolution and amplification of gauge fields during and after inflation. 
In section \ref{sec:LatticeSimulations} we describe the lattice method used and present our main numerical results, regarding the properties of the produced magnetic fields, including magnitude, helicity and correlation length, as a function of the axion-gauge coupling strength. The post-inflationary evolution of the magnetic fields is evaluated in section \ref{sec:evolution}, while our conclusions and proposed avenues for future work are presented in section  \ref{sec:conclusions}.

%%%%%%%%%%%%%%%%%%%%%%%%%%%%%%%%%%%%%%%
%%%%%%%%%%%%%%%%%%%%%%%%%%%%%%%%%%%%%%%
%%%%%%%%%%%%%%%%%%%%%%%%%%%%%%%%%%%%%%%

\section{Background and Conventions}\label{background}

We begin by defining the model, and establishing our notation and conventions. The couplings between a pseudo-scalar inflaton, $\phi$, and Abelian gauge, $A_\mu$, field have been well studied, and we gather some known results in appendix \ref{app:gaugefieldsaxion}.

In this work, we  work with the theory of an axion coupled minimally to Einstein gravity, and axially coupled to a $U(1)$ gauge field\footnote{Greek letters here and throughout denote four dimensional Lorentz indices and Roman letters from the middle of the alphabet are used to denote spatial indices.  Repeated lower spatial indices are summed using the Kronecker delta.}
\begin{align}\label{eqn:axionact}
S = \int d^4 x \sqrt{-g}\[\frac{m_{\rm pl}^2}{16 \pi}R - \frac{1}{2}\partial_\mu\phi\partial^\mu \phi - V(\phi)-\frac{1}{4}F_{\mu\nu}F^{\mu\nu} - \frac{\alpha}{4 f}\phi F_{\mu\nu}\tilde{F}^{\mu\nu}\].
\end{align}
 We work with the Friedmann-Lema\^itre-Robertson-Walker (FLRW) metric in conformal time with mostly-plus conventions, $ds^2 = -a^2(d\tau^2 - d {\bf x}^2).$
The potential, $V(\phi)$, softly breaks the axionic shift-symmetry and supports a period of slow-roll inflation \cite{Freese:1990rb, Adams:1992bn}.  For definiteness, we consider the potential for the simplest type of chaotic inflation  
$V(\phi) =  \frac{1}{2}m^2\phi^2$ \cite{Linde:1981mu}. 
The amplitude of the scalar spectrum fixes the parameters $m$ to be  $m \approx  1.06 \times 10^{-6}\,m_{\rm pl}$ \cite{Ade:2013uln}. Potentials arising in variant models, such as axion monodromy  \cite{McAllister:2008hb}, lead to very similar phenomenology \cite{Adshead:2015pva}, with possible additional phenomena, such as the emergence of oscillons \cite{Amin:2011hj,Zhou:2013tsa}. 

The field strength, $F_{\mu\nu}$, and its dual, $\tilde{F}^{\mu\nu}$, are given by the standard expressions, $F_{\mu\nu} = \partial_\mu A_\nu - \partial_{\nu}A_{\mu}$ and $\tilde{F}^{\mu\nu} = \epsilon^{\mu\nu\alpha\beta}F_{\alpha\beta}/2$,
where $\epsilon^{\mu\nu\alpha\beta}$ is the completely antisymmetric tensor and our convention is $\epsilon^{0123} = 1/\sqrt{-g}$.
The parameter $\alpha$ is a dimensionless coupling constant of order unity, and $f$ is a mass scale associated with thef pseudo-scalar (axion).  We work in so-called natural units where $\hbar = c =1$, while keeping the Planck mass, $m_{\rm pl} = 1/\sqrt{G} = 1.22 \times 10^{19}\,{\rm GeV}$. 

Gauge invariance means that we cannot simply identify $F^{\mu\nu}$ as the field strength of the electromagnetic field. Instead the axion must couple to the gauge fields corresponding to the unbroken $U(1)_Y$, and/or  $SU(2)_L$ sectors of the electro-weak theory. For simplicity we only consider coupling to $U(1)_Y$.\footnote{Coupling the axion to a non-Abelian gauge sector, like $SU(2)_L$, would also be interesting. However, the non-Abelian gauge group makes this significantly more challenging than the $U(1)$ group we consider here. We leave this case to future work.} The resulting gauge bosons will eventually become regular photons below the electroweak symmetry scale with an efficiency proportional to $\cos (\theta_W) \sim 90\%$.  Since we are eventually interested in the generation and observation of magnetic fields, we refer to the $U(1)$ hypercharge or electromagnetic field indistinguishably for the remainder of this work, keeping in mind that this is an exact correspondence only after the electroweak symmetry is broken.

%%%%%%%%%%%%%%%%%%%%%%%%%%%%%%%%%%%%%%%
%%%%%%%%%%%%%%%%%%%%%%%%%%%%%%%%%%%%%%%
%%%%%%%%%%%%%%%%%%%%%%%%%%%%%%%%%%%%%%%
%%%%%%%%%%%%%%%%%

\subsection{The electromagnetic field}
\label{sec:Gaugefieldproduction}

The generalized Maxwell equations can provide some physical insight for the helicity-dependent gauge field amplification mechanism that operates in this model. In terms of the observer's four-velocity $u^\mu$ with $u^\mu u_\mu =1$, the electric and magnetic fields can be written as 
\begin{align}
E_\mu &= F_{\mu\nu} u^\nu,\quad 
B_\mu = {1\over 2} \epsilon_{\mu\nu\alpha\beta}F^{\nu\alpha} u^\beta.
\end{align}
In an FLRW Universe the four-velocity of a co-moving observer is close to the Hubble flow, $u^\mu = a^{-1} ( 1,0,0,0)$, resulting in
\begin{align}\label{eqn:EandB}
E_\mu &= (0 , E_i ) =a^{-2} (0, \partial_\tau A_i ) , \quad 
B_\mu = (0,B_i) =a^{-2} (0, \epsilon_{ijk} \partial_j A_k ) 
\end{align}
in Coulomb gauge. Following the treatment of ref.\ \cite{Anber:2006xt}, we can write the evolution equations for the electric and magnetic fields and compare them to the usual form of Maxwell's equations. It is rather straightforward to get the generalized form of Ampere's law
\begin{align}
\partial_\tau (a^2 \vec E) = \nabla \times (a^2 \vec B) - {\alpha \over f}( \partial_\tau\phi )(a^2 \vec B) - {\alpha \over f} (\vec \nabla \phi) \times (a^2 \vec E),
\label{eq:ampere}
\end{align}
Faraday's and Gauss's laws
\begin{align}
\partial_\tau (a^2 \vec B) + \nabla \times (a^2 \vec E)=0, \quad 
\label{eq:faraday}
\vec \nabla \cdot E = -{\alpha \over f} (\vec \nabla \phi) \cdot \vec B, \quad
%\\
\vec \nabla \cdot B=0,
\end{align}
 for the electric and magnetic field.

In the limit that the inflaton field is homogeneous, ($\nabla \phi=0$), the only difference between eqs.\ \eqref{eq:ampere} and \eqref{eq:faraday} and the free Maxwell field in curved space-time is the term $(\alpha /f)( \partial_\tau\phi )(a^2 \vec B)$  that appears in Ampere's law. This takes the form of an addition to the displacement current, $\partial_\tau (a^2 \vec E)$.  This current depends on the magnetic, rather than electric, field. During the axion oscillations, $\partial_\tau \phi$ varies rapidly sourcing both electric and magnetic fields.

The second-order equation of motion for the magnetic field can be written 
\begin{align}
\left [ \partial_\tau^2 - \nabla^2 -{\alpha \over f} (\partial_\tau \phi) \nabla \times \right ] (a^2 \vec B) =0.
\label{eq:Beom}
\end{align}
Note that the term involving the curl distinguishes left-handed from right-handed polarized modes. The helicity-dependence is immediately apparent at the level of the magnetic field equation.

As can be immediately seen from eq.\ \eqref{eq:Beom}, in the absence of strong interaction with an inflaton condensate, the B-field will decay as $B_i \sim a^{-2}$. In a conducting plasma, as we expect the Universe to behave after reheating, the electric field will be quickly damped.

\subsection{Electro-magnetic power spectra}
\label{sec:powerspectra}

In order to study the evolution of the magnetic fields in the primordial plasma and beyond, we define the magnetic field strength and helicity spectra \cite{Durrer:2010mq,Jain:2012jy,Durrer:2013pga,Campanelli:2007tc}. The magnetic fields generated during inflation are statistically homogenous and isotropic, therefore its two-point correlation in position space $\langle B_i(x) B_i(y) \rangle$ is only a function of the distance $|\vec x-\vec y|$. It is more convenient to work in Fourier space, where the two-point function can be written as
\begin{align}
a^4\langle B_i(\vec k,t) B_j^* (\vec q,t) \rangle = {(2\pi)^3 \over 2 } \delta(\vec k - \vec q) \left [ (\delta_{ij} - \hat k_i \hat k_j )P_S(k,t) - i \epsilon_{ijl} \hat k_l P_A (k,t) \right ],
\end{align}
where $\left <... \right > $ is an ensemble average, which can be thought of as a spatial average over many patches due to the ergodic theorem \cite{Durrer:2013pga}.  The functions $P_S$ and $P_A$ are the symmetric and anti-symmetric components of the power spectrum, related to the energy density and helicity density of the magnetic field respectively.
We can write the power spectrum as a function of the helicity components of the electromagnetic vector potential, 
\begin{align}
P_{S} (k,t) &= k^2 \left ( |A_+(k,t) |^2 + |A_-(k,t) |^2    \right )
\\
P_{A} (k,t) &= k^2 \left ( |A_+(k,t) |^2 - |A_-(k,t) |^2    \right ) \, ,
\label{eq:PsPa}
\end{align}
where the usual decomposition of the momentum-space gauge field potential $A_\mu(k,t)$ into helicity modes $A_\pm(k,t)$ is described in appendix \ref{app:gaugefieldsaxion}.
The magnetic energy can be written as an integral of the magnetic energy density in position or momentum space,
\begin{align}
E_B(t) &= {1\over V} \int_V d^3x \, { \langle B^2 \rangle \over 2}= \int_0^\infty dk \, {\cal E}_B(k,t),
\label{eq:EBdef}
\end{align}
and correspondingly for the magnetic helicity ,
\begin{align}
H_B(t) &= {1\over V} \int_V d^3x \, \langle A\cdot B \rangle = \int_0^\infty dk \, {\cal H}_B(k,t) ,
\label{eq:HBdef}
\end{align}
where 
\begin{align}
{\cal E}_B &= k^2 {P_S\over (2\pi)^2}\quad \text{and} \quad 
{\cal H}_B = k {P_A\over 2\pi^2},
\end{align}
are the relations between the energy and magnetic helicity densities and the symmetric and antisymmetric parts of the power spectrum respectively.

A maximally helical field occurs for $P_S= P_A$ and a completely non-helical one for $P_A=0$. Note that  eq.\ \eqref{eq:PsPa} implies that any magnetic field configuration must satisfy 
\begin{align}
|{\cal H}_B| \le 2k^{-1} {\cal E}_B
\label{eq:realiz}
\end{align}
with the equality holding for maximally helical fields. 

Previous studies of gauge-field production from axion inflation concluded that an effectively maximally helical power spectrum is produced, since the mode that is amplified during inflation grows by several orders of magnitude more than the mode that is only amplified during preheating. While this is an excellent approximation for low axion-gauge couplings, we have shown in ref.\ \cite{Adshead:2015pva} that this changes drastically in the medium-to-large coupling regime, where re-scattering effects between the gauge and axion modes are important. In this case the subdominant helicity  is produced through scattering of the dominant mode off the axion. For large wave-numbers, this leads to a largely non-helical field, while for smaller wave-numbers (super-horizon modes) a significant net polarization remains. Due to the scale-dependence of the helicity fraction, we use an integral form of eq.\  \eqref{eq:realiz} as in ref.\ \cite{Campanelli:2007tc}. We first define the magnetic comoving correlation length $\xi_B$ as
\begin{align}
\xi_B = {1\over E_B} \int_0 ^\infty {dk \over k} {\cal E_B} \, ,
\label{eq:xi_B}
\end{align}
which is also the scale at which eddies develop in  MHD turbulence. Using the definition in eq.\ \eqref{eq:xi_B}, eq.\ \eqref{eq:realiz} can be integrated to give
\begin{align}
|H_B| \le 2 \xi_B E_B .
\label{eq:realizInt}
\end{align}
We demonstrate that for large values of the axial coupling the characterization of the resulting field's helicity is rather non-trivial and we need to make further assumptions to use the known results from the literature regarding the evolution of magnetic fields in a plasma. 

We define the physical intensity and physical correlation length of the  magnetic field as in refs.\ \cite{Durrer:2013pga,Caprini:2014mja,Fujita:2015iga},
\begin{align}
B^2_{\rm phys} &= {1\over a^4} \int { d^3k \over (2\pi)^3 } k^2 (|A_k^+|^2 + |A_k^-|^2 )  =  {1\over 2\pi^2 a^4} \int dk k^4 (|A_k^+|^2 + |A_k^-|^2 )
\label{eq:B2phys}
\\
\lambda_{\rm phys} &= {1\over B^2_{\rm phys} a^3} \int { d^3k \over (2\pi)^3 } k^2 {2\pi \over k} (|A_k^+|^2 + |A_k^-|^2 )  = 2\pi a  {\int dk k^3 (|A_k^+|^2 + |A_k^-|^2 ) \over \int dk k^4 (|A_k^+|^2 + |A_k^-|^2 )}.
\label{eq:lambdaphys}
\end{align}
The physical quantities for the magnetic field strength and correlation length are trivially related to their comoving counterparts by $\lambda_{\rm phys} = 2\pi a \xi_B$ and $B^2_{\rm phys}  =a^{-4} E_B$. We further define the energy density in the hyper-electromagnetic field as
\begin{align}
\rho_{EM} = {1\over 2} \langle  E^2 + B^2 \rangle \,  ,
\end{align}
which we use to calculate the efficiency of the energy transfer from the axion condensate to the gauge fields. Finally we define the energy density in the fluctuations of the axion as 
\begin{align}
\rho_{\delta\phi} = {1\over 2} m^2 \langle \delta \phi^2 \rangle \, ,
\end{align}
where $\langle \delta \phi^2 \rangle$ is the variance of the axion-inflaton perturbations. 

%%%%%%%%%%%%%%%%%%%%%%%%%%%%%%%%%%%%%%
%%%%%%%%%%%%%%%%%%%%%%%%%%%%%%%%%%%%%%

%%%%%%%%%%%%%%%%%%%%%%%%%%%%%%%%%%%%
%%%%%%%%%%%%%%%%%%%%%%%%%%%%%%%%%%%%

\section{Lattice Simulations}
\label{sec:LatticeSimulations}

For small values of the coupling, we can calculate the generated gauge fields using the linear (no-backreaction) approximation, which is defined as follows.  The equations of motion for the inflaton field, eq.\ \eqref{axioneom}, and Hubble parameter, eq.\ \eqref{ffriedman}, are solved numerically by neglecting the effect of the gauge fields. The resulting $a(t)$ and $\phi(t)$ are then used to numerically solve the equations of motion for the gauge field mode by mode, as in eq.\ \eqref{eqn:kspacegfeqn}.

Since our interest in this work focuses on the regime of large coupling and strong backreaction, we need to go beyond the linear regime and accurately model the non-linear processes. For that we need to use lattice simulations. In this section we start by briefly describing our numerical methods and providing tests for the accuracy of the initialization procedure we employ. Section \ref{sec:simulationresults} contains a presentation and extensive discussion of the results in all allowed regimes of the axion-gauge coupling strength.

\subsection{Initial conditions and initialization}

We use  {\sc GABE} \cite{GABE} and the strategies described in refs.\ \cite{Deskins:2013dwa, Adshead:2015pva} to evolve the gauge and axion fields together with the background spacetime.  The axion and the gauge field are defined on a discrete lattice (grid) with $256^3$ points and a second-order Runge-Kutta integration method solves eqs.\ \eqref{axioneom}, \eqref{gaugeeom1} and \eqref{gaugeeom2} alongside the self consistent expansion of space-time eq.\ \eqref{ffriedman}. 

We work in Lorenz gauge, $\partial^\mu A_\mu = 0$, to evolve the gauge fields. In this gauge, the Gauss law constraint becomes a dynamical equation for $A_0$ which we solve in parallel with eq.\ \eqref{gaugeeom2}. As explained in ref.\ \cite{Adshead:2015pva}, although we initialize our fields using solutions of the linear equations of motion in Coulomb gauge, these gauge fields are equivalent to the fields in Lorenz gauge.  Unless otherwise noted, all simulations use a box-size that is $L=15\,{m^{-1}}$ at the end of inflation and are run using the parallel processing standard {\sc OpenMP} \cite{dagum1998openmp}, generally with 12 threads.  

Because  all fields are not in their vacuum states at the end of inflation due to the tachyonic enhancement of the gauge modes near their horizon crossing, we begin our simulations two $e$-foldings before inflation ends. At this point, almost all modes of interest for the reheating phase are within the horizon, and thus their initial states at the beginning of reheating are dynamically generated.  While the origin of these inhomogeneities are quantum mechanical, in our simulations they are treated classically.  This is consistent since, while they do contribute to the energy density of the simulation, this contribution is small compared to the homogeneous background until they are amplified or become super-horizon, in which cases they are in the classical regime.

To generate the initial spectra two $e$-folds before the end of inflation, we first determine the value of the homogeneous field and its derivative by numerically evolving the system of eqs.\ \eqref{eqn:backreactfried}-\eqref{eqn:backreactKG} together with the approximations of eq.\ \eqref{eqn:GFenergy}.  At this point, the box-size is $L_0 = L e^{-2} \approx 2 \,m^{-1}$, just larger than the Hubble scale, $H^{-1}  = \sqrt{3/8\pi}\rho^{-1/2} \approx 1.2\, m^{-1}$, where the final approximation varies slightly for each coupling. We then initialize the power in the $A_{\pm}$ modes using a two step process.  First we numerically evolving eq.\ \eqref{eqn:kspacegfeqn} for a dense set of physical wavenumbers, tracking each mode from when it was well inside the horizon ($\tau \rightarrow -\infty$) until two e-foldings before the end of inflation.  Then we use this as a template power spectrum that defines the distributions from which we initialize each independent mode.\footnote{An independent prescription for the initial conditions for lattice simulations of gauge preheating was recently presented in ref.\ \cite{Lozanov:2016pac}.} 

After setting the initial spectrum of $A^\pm_{{\bf k}}$ in momentum space, we construct the Fourier space gauge fields
\begin{equation}
\vec{A}_{\bf k}  = \vec{\varepsilon}_+({\bf k}) A^+_k  + \vec{\varepsilon}_-({\bf k}) A^-_{k},
\end{equation}
where $\vec{\varepsilon}_\pm({\bf k})$ are the polarization vectors defined in eq.\ \eqref{eqn:polvecs}. We then (inverse) Fourier transform them into configuration space. These relations set only the spatial components of the gauge field, $\vec{A}({\bf x}, \tau = 0)$, on the initial surface.  Since we are numerically tracking the values of the full four-potential, $A^\mu$, we must check to make sure the Lorenz gauge condition, $\partial^\mu A_\mu = 0$, is obeyed in configuration space, as required by our equations of motion. The definition of the polarizations, eq.\ \eqref{eqn:polvecs}, requires that $\dot{A}^0 = 0$ ($A^0 = {\rm constant}$) on the initial slice.  Therefore any choice of $\vec{A}_\pm$ (with the choice $A_0=0$) obeys the gauge condition.  

While the gauge fields during inflation are accurately described by the solutions of their linearized equations, the fluctuations of $\phi$ are modified from their linear form by non-linear interactions with the gauge fields. This makes their accurate solution difficult. However, the backreaction of the gauge modes does not become important until near horizon crossing for a given wavenumber. Therefore, we initialize the inflaton fluctuations in the Bunch-Davies vacuum, $\left\langle\phi_k^2\right\rangle = 1/\sqrt{2\omega}$, for our simulations. This is an excellent approximation due to the fact that almost all of our modes are sub-horizon, introducing only a small error for the modes that are near to the horizon at the initial time. Using this procedure, the modes that leave the horizon during the final two e-foldings of inflation are generated self-consistently and the resulting spectrum of perturbations for $\phi$ is consistent with our equations of motion. We justify this approximation in the next section.

\begin{figure}[t]
\centering
\includegraphics[width = \textwidth]{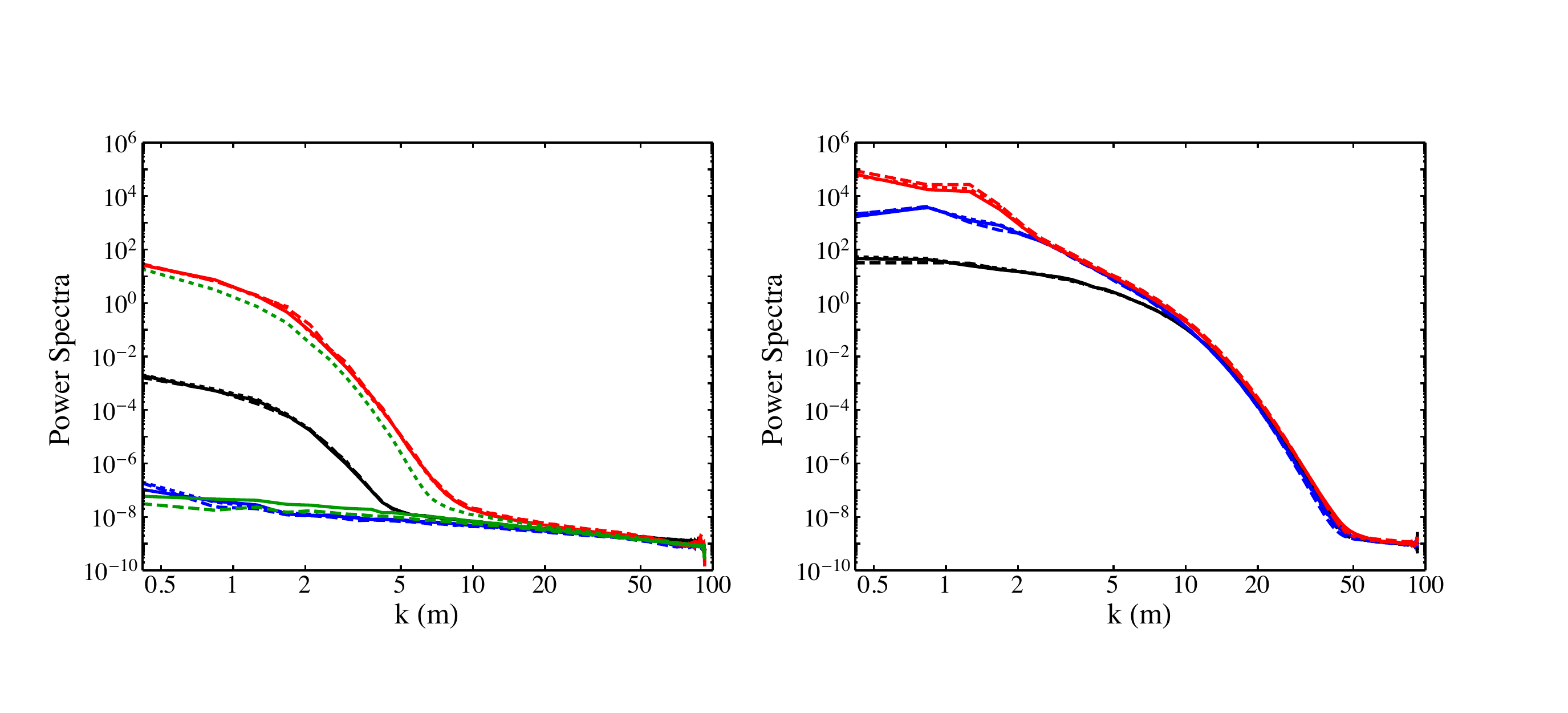} 
\caption{  Resulting power spectra for $A^-$ (red), $A^+$(blue) and $\delta \phi$ (black) at the end of inflation (left panel) and at $N=2$ efolds after the end of inflation (right panel). The dotted, solid, and dashed lines correspond to simulations initialized at $N=-2.1$, $N=-2$, and $N=-1.8$ respectively. The green lines correspond to the initial conditions for $\delta \phi$ (solid), $A^-$ (dotted), and $A^+$ (dashed) at $N=0$. }
\label{fig:varyNstart}
\end{figure}

As a check on the validity of our initial conditions generated as described above, we perform the simulation by keeping the coupling fixed and varying the starting time to ensure that the results are unchanged. In figure \ref{fig:varyNstart} we compare the final spectra for the gauge fields and the axion fluctuations that arise from choosing $\alpha / f = 60 m_{\rm Pl}^{-1}$ and initializing the simulation at $N=-2.1$, $N=-2$, and $N=-1.8$. Between $N=-2.1$ and $N=-1.8$ the Universe would have expanded by a factor of about $2.5$. We can see no appreciable difference between these three runs hence initializing our code even earlier will only use more computational resources without increasing the accuracy of our results. The small differences that are still visible are an artifact of using a finite grid and sampling the modes from a distribution with random amplitudes and phases. We present a short discussion of the systematic sampling error in appendix \ref{app:sampling}. 

As a demonstration of the benefit of starting the calculation during inflation, we overlay the initial conditions that our prescription would provide for a simulation starting at $N=0$. We see that there is visible difference in the spectrum of the dominant helicity mode, with the initial conditions at $N=0$ underestimating the amplitude by an ${\cal O}(1)$ factor. The largest difference appears in the spectrum of the axion fluctuations, which we always initialize to follow a Bunch-Davies distribution. However, for large couplings the generation of axion perturbations through the gauge fields is significant, even during inflation. Even with these differences between the spectra, most of the energy density at the end of inflation is carried by the dominant helicity mode, hence the final spectra will be qualitatively similar even for a simulation starting at $N=0$. The main qualitative difference will be the time evolution of quantities like $|\delta \phi|^2$, which can have significant consequences for the formation of oscillons in axion monodromy models \cite{Adshead:2015pva}.

\subsubsection{Initial conditions for $\delta\phi$}
A further test of our method of initializing the lattice code is the initial conditions of the inflaton perturbations $\delta\phi$. Once gauge fields are produced, they backreact on the inflaton,  generating an additional, non-homogeneous part of the $\delta\phi$ spectrum.
While  the formalism of eq.\ \eqref{eq:deltaphi_spectrum} gives an estimate of the backreacted power spectrum, we make one further approximation, as a means to quantify the maximum deviation of $\delta \phi_k$ from the Bunch-Davies spectrum. Following the calculations given in refs.\ \cite{Barnaby:2011vw, Barnaby:2010vf} and further applied and summarized in \cite{Linde:2012bt,Bugaev:2013fya}, the backreaction-seeded part of power spectrum can be approximated by 
\begin{align}
\left < \delta \phi_{\rm BR} ^2 \right > \approx {\alpha^2 \over f^2} {\left < \vec E \cdot \vec B \right > ^2 \over 9 \beta^2 H^4}, \quad \beta \equiv 1 - {2\pi \xi} {\alpha \over f} { \left < \vec E \cdot \vec B \right >  \over 3 H \dot \phi},
\end{align}
for modes that are close to horizon-crossing.

On the one hand, using the approximate form for the power spectrum of the amplified gauge field during inflation eq.\ \eqref{eq:approximateAk}  along with the numerical solutions arising from eqs.\ \eqref{eqn:backreactfried}-\eqref{eqn:backreactKG},  for $\alpha/f = 60\, m_{\rm Pl}^{-1}$  we find $ \left < \delta \phi_{\rm BR} ^2 \right > \approx 5 \times 10^{-11} \,m_{\rm Pl}^2$ evaluated $2$ e-folds before the end of inflation. On the other hand, the power spectrum of a free scalar field is $ \left < \delta \phi_{\rm free} ^2 \right > = H^2 / (2\pi) \approx 4  \times 10^{-11} \,m_{\rm Pl}^2$. The two spectra are at most comparable,  even for modes at horizon crossing, where the backreaction is expected to be the largest. For the main set of simulations performed for the present paper, where the starting time was taken to be $2$ e-folds before the end of inflation, the mode that exits the horizon at that time $k=aH$ can be calculated to correspond to a comoving wavenumber of $k\sim 0.1\, m$. The minimum comoving wavenumber captured by our simulation is $k_{\rm min} = 2\pi /L$ which turns out to be $k_{\rm min} \simeq 0.4\,m$. Hence all modes that are simulated are well inside the horizon and thus are expected to closely follow the BD distribution.

\begin{figure}[t]
\centering
 \includegraphics[width = \textwidth]{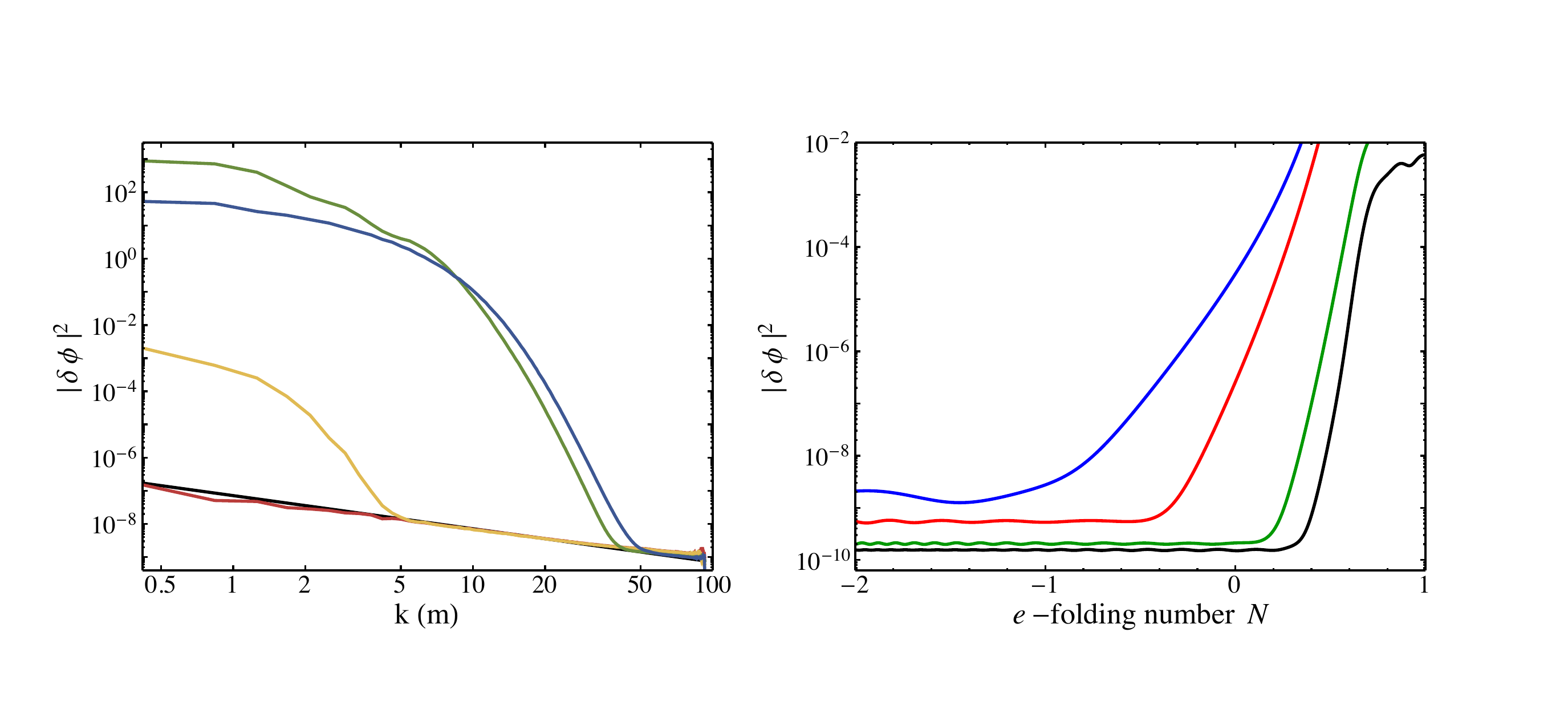}
 \caption{ 
The left panel shows the power spectrum of the axion fluctuations $\delta \phi$ for $N=-2$, $N=-1$, $N=0$, $N=1$, and $N=2$ for $\alpha /f=60\,m_{\rm Pl}^{-1}$ color-coded in a rainbow scale with $N=-1$ corresponding to red and $N=2$ corresponding to blue. The initial Bunch-Davies spectrum at $N=2$ is plotted in black.   The right panel shows the evolution of wavenumbers $k = 2\pi/L$, $k=10\pi/L$, $k=30\pi/L$, and $k=40\pi/L$ (blue, red, green and black respectively) as a function of the number of e-folds. We can see the lowest $k$ mode is excited around $N=-1$, which validates our choice of initial conditions.  
}
\label{fig:BD_deltaphi}
\end{figure}

As a numerical test of our conclusion that the Bunch-Davies initial conditions for the initial inflaton perturbation spectrum are enough for the purposes of our simulation, we plot the spectrum of $|\delta \phi|^2$ for different times at figure \ref{fig:BD_deltaphi}. We see that the backreaction during inflation is small enough that no appreciable excitation of inflaton modes is seen between $N=-2$ and $N=-1$ for the simulated wave-numbers of interest. Figure \ref{fig:BD_deltaphi} also shows the time-evolution of some $k$-modes of $|\delta\phi|^2$. An oscillatory behavior can be seen for low wavenumbers and early times. This is an artifact of the finite number of wavenumbers in each bin, as explained in appendix \ref{app:sampling}.

%%%%%%%%%%%%%%%%%%%%%%%%%%%%%%%%%%%%
%%%%%%%%%%%%%%%%%%%%%%%%%%%%%%%%%%%%
%%%%%%%%%%%%%%%%%%%%%%%%%%%%%%%%%%%%

%%%%%%%%%%%%%%%%%%%%%%%%%%%%%%%%%%%%%%%%%%%%%%%%%%%
%%%%%%%%%%%%%%%%%%%%%%%%%%%%%%%%%%%%%%%%%%%%%%%%%%%
%%%%%%%%%%%%%%%%%%%%%%%%%%%%%%%%%%%%%%%%%%%%%%%%%%%
%%%%%%%%%%%%%%%%%%%%%%%%%%%%%%%%%%%%%%%%%%%%%%%%%%%

\subsection{Hypermagnetic field simulations}
\label{sec:simulationresults}

\begin{figure}[t]
\centering
\includegraphics[width=\textwidth]{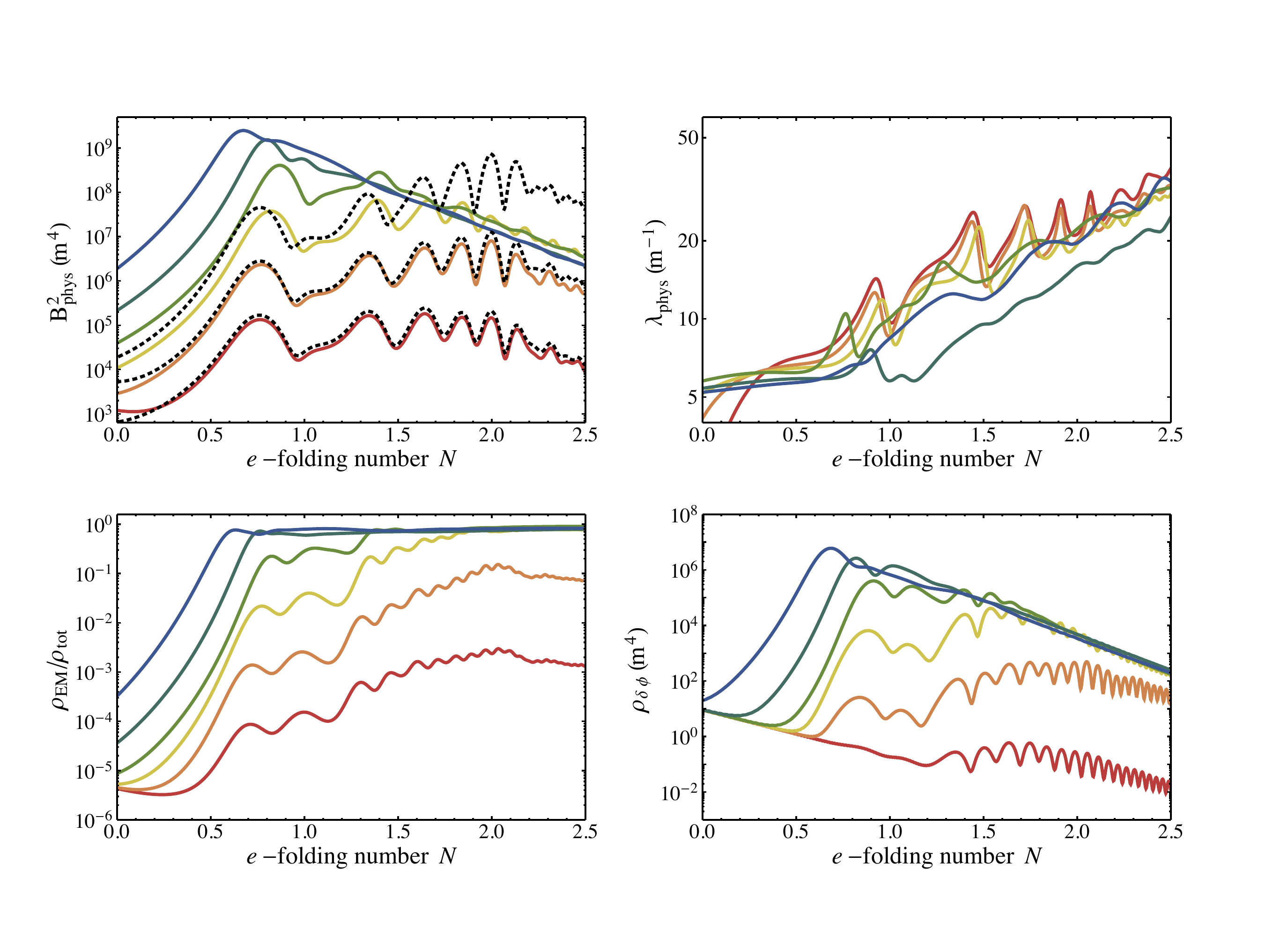}
\caption{ Clockwise from the top: The magnetic field intensity, the correlation length, the energy density of inflaton fluctuations, and the total energy fraction in the gauge field. All are plotted as a function of the number of e-folds after the end of inflation for several values of the axion-gauge coupling from $\alpha/f = 35 \,m_{\rm Pl}^{-1}$ (red) to $\alpha/f = 60\, m_{\rm Pl}^{-1}$ (purple) in increments of $\alpha/f = 5\, m_{\rm Pl}^{-1}$, color-coded along the rainbow spectrum. The dotted black curves of $B^2_{\rm phys}$ show the results of the the no-backreaction calculation.
}
\label{fig:parameterscan}
\end{figure}

We begin by presenting the global results of the simulations of the hypermagnetic field evolution across a range couplings from $\alpha / f = 35 \,m_{\rm Pl}^{-1}$ to $\alpha / f = 60 \, m_{\rm Pl}^{-1}$, safely below the limit of $\alpha /f \le 110\, m_{\rm Pl}^{-1}$, set by constraints on the production of primordial black holes \cite{Linde:2012bt,Bugaev:2013fya}.
  In figure \ref{fig:parameterscan} we show the evolution of the magnetic field, the correlation length, the power in axion perturbations and the energy fraction in the electromagnetic field.  

We find that the correlation length is almost constant, apart from the fine oscillatory features, among many runs with varying coupling strength. This behavior is due to the fact that the correlation length is set by the horizon size at the end of inflation which does not strongly depend on the coupling. As discussed in ref.\ \cite{Adshead:2015pva}, for $\alpha / f > 50\, m_{\rm Pl}^{-1}$ the transfer of energy from the axion condensate to the gauge fields occurs within a single  oscillation of the axion background. In fact by further increasing the axion-gauge coupling, the gauge fields produced during inflation generate an extra drag force that acts onto the inflaton, eventually making the system over-damped, so that the inflaton condensate does not even complete one single oscillation. The behavior of the magnetic field and the variance of the axion perturbations are similar, in the fact that they increase with increasing coupling and effectively saturate for $\alpha/f \simeq 60 \, m_{\rm Pl}^{-1}$. 
We now consider each range of couplings separately and analyze the behavior of the gauge fields in each one, so us to understand the underlying physical properties. Where applicable, we compare the lattice results to the no-backreaction calculations described in the beginning of section \ref{sec:LatticeSimulations}.

\subsubsection{Small Coupling}
\label{sec:SmallCoupling}

We start at small couplings, where we can test the results of our lattice code against the linear calculation. In this region, backreaction is small, which means that the linear theory is an accurate approximation to the full evolution. We characterize the size of the coupling by the resulting energy density stored in the $U(1)$ field at the end of the preheating stage, where the transfer of energy from the inflaton condensate to the gauge fields has effectively ceased, or by the amount of backreaction of the gauge fields on the inflaton. These two indicators are essentially equivalent.  As an example, we choose to analyze a coupling of ${\alpha / f} = 35 \,m_{\rm Pl}^{-1}$. As  in reference \cite{Adshead:2015pva}, this coupling results in $\rho_{\rm EM} / \rho_{\rm tot} = {\cal O}(10^{-3})$, not enough to preheat the Universe.

\begin{figure}[t]
\centering
\includegraphics[width=\textwidth]{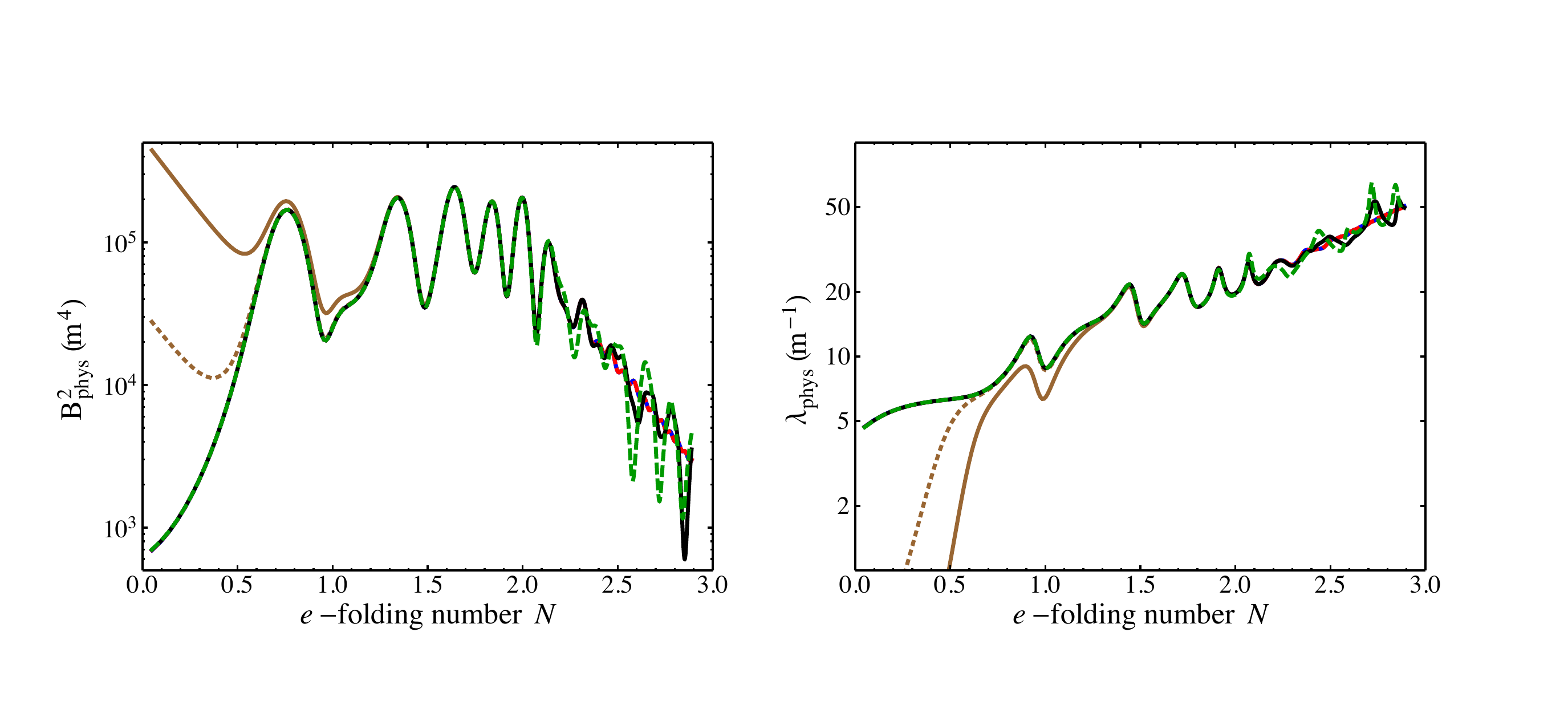}
\caption{ The dependence of the physical magnetic field strength $B^2_{\rm phy}$ and the physical correlation length $\lambda_{\rm phy}$ as a function of the wave-number discretization. The brown and brown-dotted curves correspond to integrating up to a maximum comoving wavenumber of $80\,m$ and $20\,m$ respectively. All others correspond to integrating up to $k=10\,m$ for $\Delta k = \pi / 30\,m$, $\Delta k = 2\pi / 30\,m$, $\Delta k =3\pi / 30\,m$, and  $\Delta k =4\pi / 30\,m$ (blue, red-dashed, black and green-dashed respectively)}
\label{fig:Deltak}
\end{figure}

One challenge that presents itself when using lattice methods is that of scales; it is computationally expensive to introduce more lattice points, especially in three dimensions, as the simulation progresses and the physical size of the box expands. The number of momentum modes we can include is finite. In this specific set of simulations we follow the behavior of $128^3/2$ modes. The output data are presented by combining the modes into bins separated by $\Delta k = {2\pi / (15,m^{-1})} \simeq 0.4 \, m$, which is also the shortest comoving wave-number that we can probe. By performing a no-backreaction calculation using {\sc Mathematica}, we can examine the effect of using various numbers of grid-points, or, equivalently, various $\Delta k$, for the same maximum simulated wavenumber $k_{\rm max}$ on the calculated quantities, such as the magnetic field strength and correlation length. In figure \ref{fig:Deltak} we present the result of varying the number of $k$ modes. Reducing the number of $k$ modes leads to larger oscillations at late times, where the amplification of the gauge fields has essentially ceased and the magnetic field is simply red-shifting due to the expansion of the Universe. These oscillations are due to the fine band structure in the power spectrum of the gauge fields at these late times. A sparse $k$-grid can therefore lead to significant sampling errors by randomly sampling or missing each band. However, a simple time-averaging can still bring out the underlying red-shifting behavior. 

Furthermore, at early times, the numerical results for the magnetic field and correlation length obtained from the integration of the expressions given in eq.\ \eqref{eq:B2phys} and eq.\ \eqref{eq:lambdaphys} are somewhat dependent on the upper limit of integration, as seen in figure \ref{fig:Deltak}.  This is indicative of a renormalization issue that we have not addressed. If one would integrate the mode amplitude in an infinite $k$ interval, the Bunch Davies contribution would cause the integral to diverge. In any reasonable finite range of wavenumbers, once the tachyonic resonance sets in, the modes that are amplified dominate the energy density, and correspondingly the magnetic field intensity. Renormalization is not be required, unless the considered range of wavenumbers is exponentially larger than the range of amplified wavenumbers.  We return to this issue in section \ref{sec:LargeCoupling}. 

\begin{figure}[t]
\centering
\includegraphics[width = \textwidth]{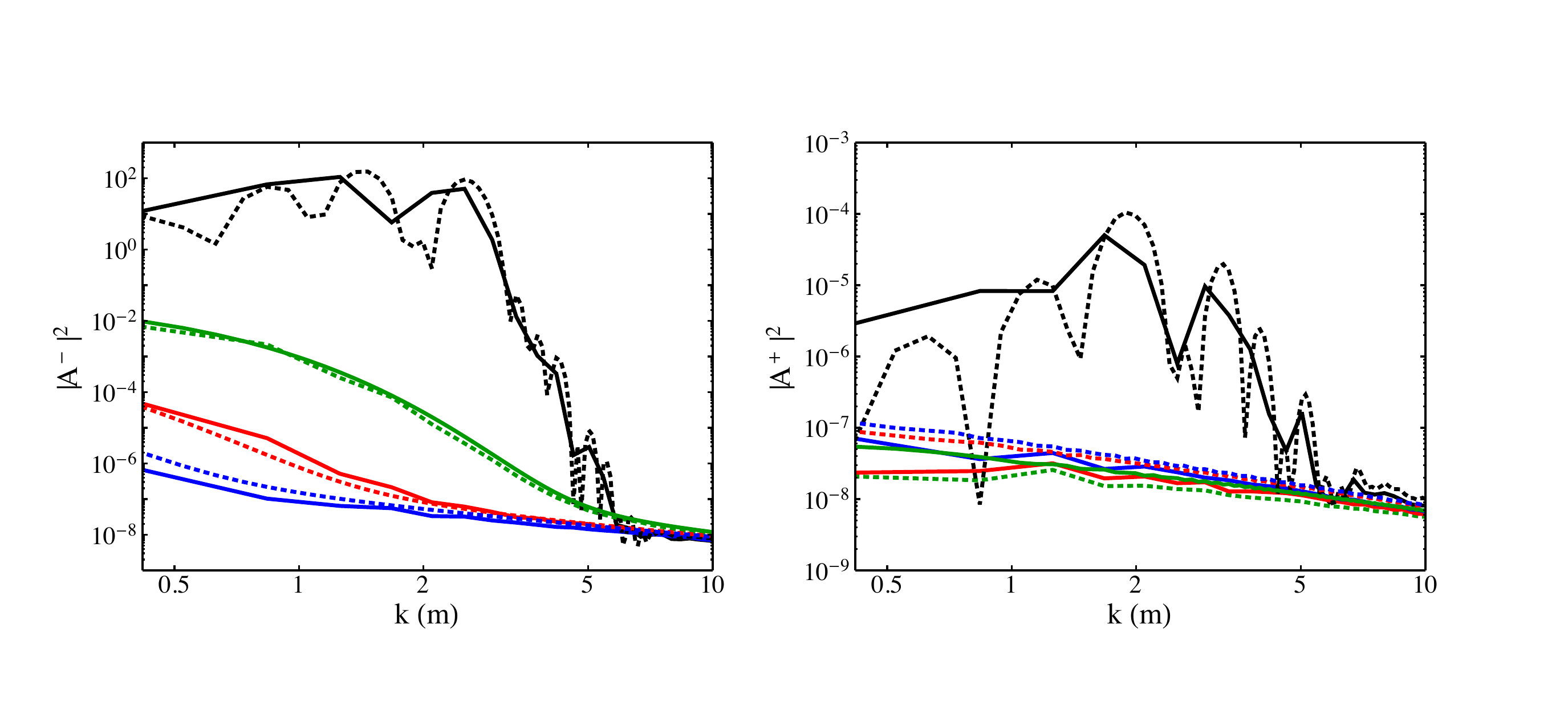}
\caption{ The spectra for $\alpha /f = 35 \, m_{\rm Pl}^{-1}$ for the negative (left panel) and positive (right panel) helicity modes for $N=-2$, $N=-1$, $N=0$, and $N=2$
 e-folds (blue, red, green and black respectively). The dotted lines correspond to the no-backreaction calculation and the solid lines show the lattice results. }
\label{fig:spectra35}
\end{figure}

The power spectra for the two gauge field polarization modes calculated through the no-backreaction approximation and using the full lattice code are in excellent agreement, as shown in figure \ref{fig:spectra35}.

\subsubsection{Moderate Coupling}
\label{sec:ModerateCoupling}
For moderate values of the coupling, we distinguish between complete preheating, where the Universe is radiation dominated at the end of preheating,  and incomplete preheating, where the Universe is matter dominated at the end of preheating.  For a coupling of ${\alpha / f} = 40 \,m_{\rm Pl}^{-1}$ --- which is approximately the same as the characteristic value of ${\alpha / f} = 8 \,m_{\rm Pl}^{-1}$ presented in \cite{Fujita:2015iga} ---  the final energy density in the gauge fields is roughly $10\%$ of the total energy density of the Universe. While significant levels of gauge fields are produced, adding up to an an ${\cal O}(1)$ fraction of the total energy density, the resulting backreaction is small, and as a result the linear analysis gives a very similar result to our full lattice simulation. Figure \ref{fig:parameterscan} shows the results of our no-backreaction calculation and of the lattice simulation. 

However, deviations between the linear (no-backreaction) results and the lattice results are evident in the spectra for $|A^{\pm}(k)|^2$. First,  the predictions for subdominant mode are very different. This is easily understood. Based on the no-backreaction approximation, the positive helicity mode is only amplified after the end of inflation, when the inflaton velocity $\dot \phi$ changes sign. As explained in \cite{Adshead:2015pva}, this leads to a vast difference --- by several orders of magnitude --- of the amplification of the two helicities. The helicity that is excited during inflation attains a much larger amplitude relative to the one that is only excited after inflation has ended. However, photons of the dominant helicity mode can re-scatter off the axion and generate photons of the opposite helicity.\footnote{In this context the term photon is used describe the quanta of the $U(1)_Y$ field.} While re-scattering is certainly active for these moderate couplings, the resulting helicities still differ by orders of magnitude and the observables like the magnetic field amplitude and the electromagnetic energy density are dominated by a single helicity. Second, the lattice and no-backreaction results differ even for the dominant helicity mode. Specifically, the maximum excited wavenumber $k$ is increased at late times compared to the no-backreaction case. This means that re-scattering effects allow for the population of wavenumbers that would be unaccessible by means of tachyonic or parametric resonance. Wavenumbers smaller than this cutoff value behave similarly for the lattice and the no-backreaction calculations and it is these wavenumbers that are most amplified, hence the resulting magnetic field strength, $B^2$, shows no visible difference between the lattice and the no-backreaction calculations, as shown in figure \ref{fig:parameterscan}.  We return to this effect in the next section, section \ref{sec:LargeCoupling}, where we consider larger couplings.

\begin{figure}[t]
\centering
\includegraphics[width = \textwidth]{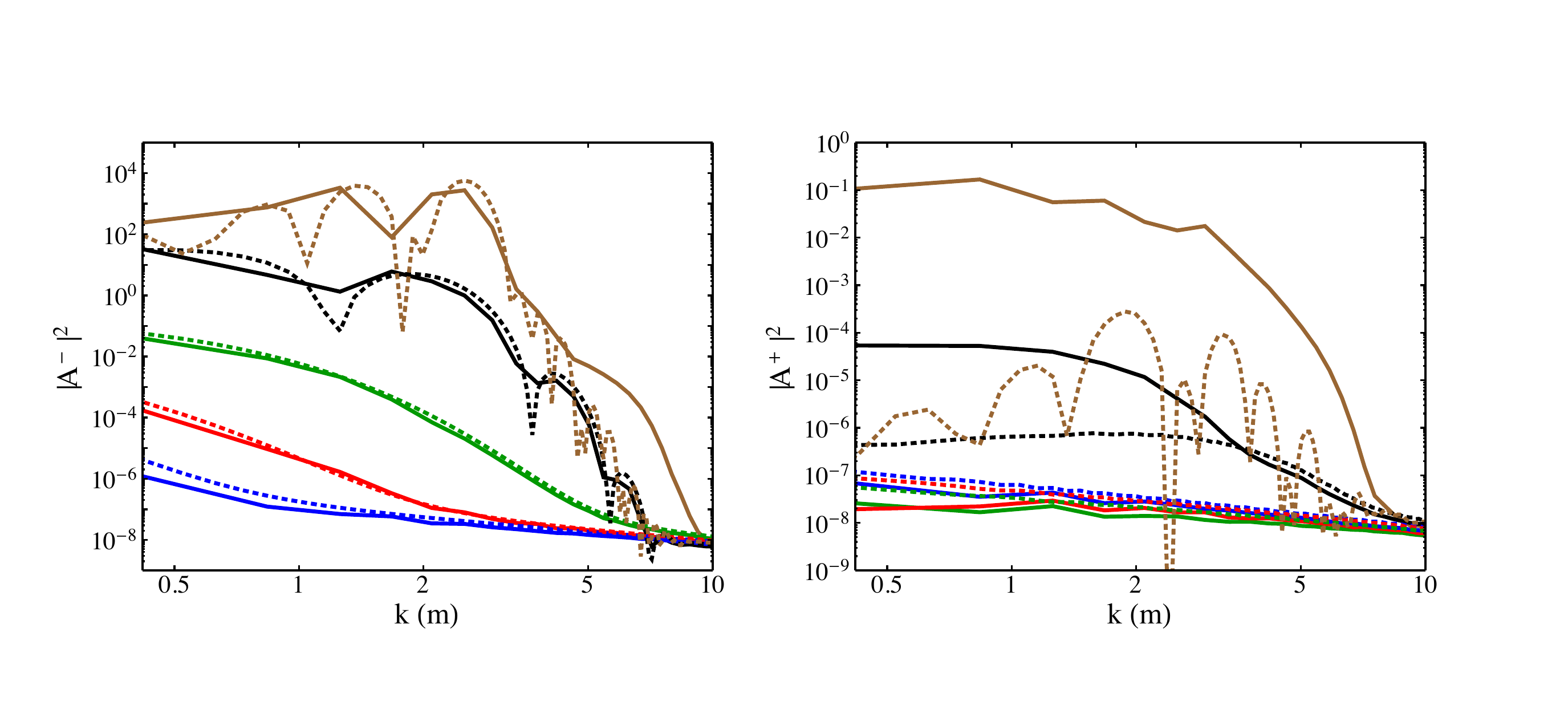}
\caption{ The spectra for $\alpha /f = 40 \,m^{-1}_{\rm Pl}$ for the negative (left panel) and positive (right panel) helicity modes for $N=-2$, $N=-1$, $N=0$, $N=1$, and $N=2$
 e-folds (blue, red, green, black and brown respectively). The dotted lines correspond to the no-backreaction calculation and the solid lines show the lattice results. }
\label{fig:spectra40}
\end{figure}

At a coupling of ${\alpha / f} = 45 \,m_{\rm Pl}^{-1}$, all of the energy density of the Universe ends up in gauge fields at the end of the preheating stage.  This is not yet in the most efficient regime, however, as it takes several oscillations of the background axion field about the minimum of its potential for this transfer of energy to occur, as shown in ref.\ \cite{Adshead:2015pva} and figure  \ref{fig:parameterscan}. In this case an interesting departure starts to show between the no-backreaction and lattice results for the intensity of the hyper-magnetic field $B^2$. First of all, there is a time-offset in the position of the first two peaks, which are otherwise very similar. This occurs because the extra friction from the gauge fields produced during inflation on the axion field causes inflation to end at a slightly different value of the background axion field compared to the no-backreaction case, as explained and calculated in ref.\ \cite{Adshead:2015pva}. Second, the most significant departure is the late-time behavior. We see that after the first two peaks the no-backreaction calculation predicts significantly higher production of hyper-magnetic fields than the lattice simulation. This is to be expected, since in this case the axion condensate is completely dissolved by transferring all its energy into the gauge fields, which is captured by the lattice simulation. From this point onward the backreaction is strong enough to invalidate the no-backreaction approximation.
%%%%%%%%%%%%%%%%%%%%%%%%%%%%%%%%%%%%%%%%%%%%%%%%%%%
%%%%%%%%%%%%%%%%%%%%%%%%%%%%%%%%%%%%%%%%%%%%%%%%%%%

\subsubsection{Large Coupling}
\label{sec:LargeCoupling}

\begin{figure}[t]
%\centering
\includegraphics[width=\textwidth]{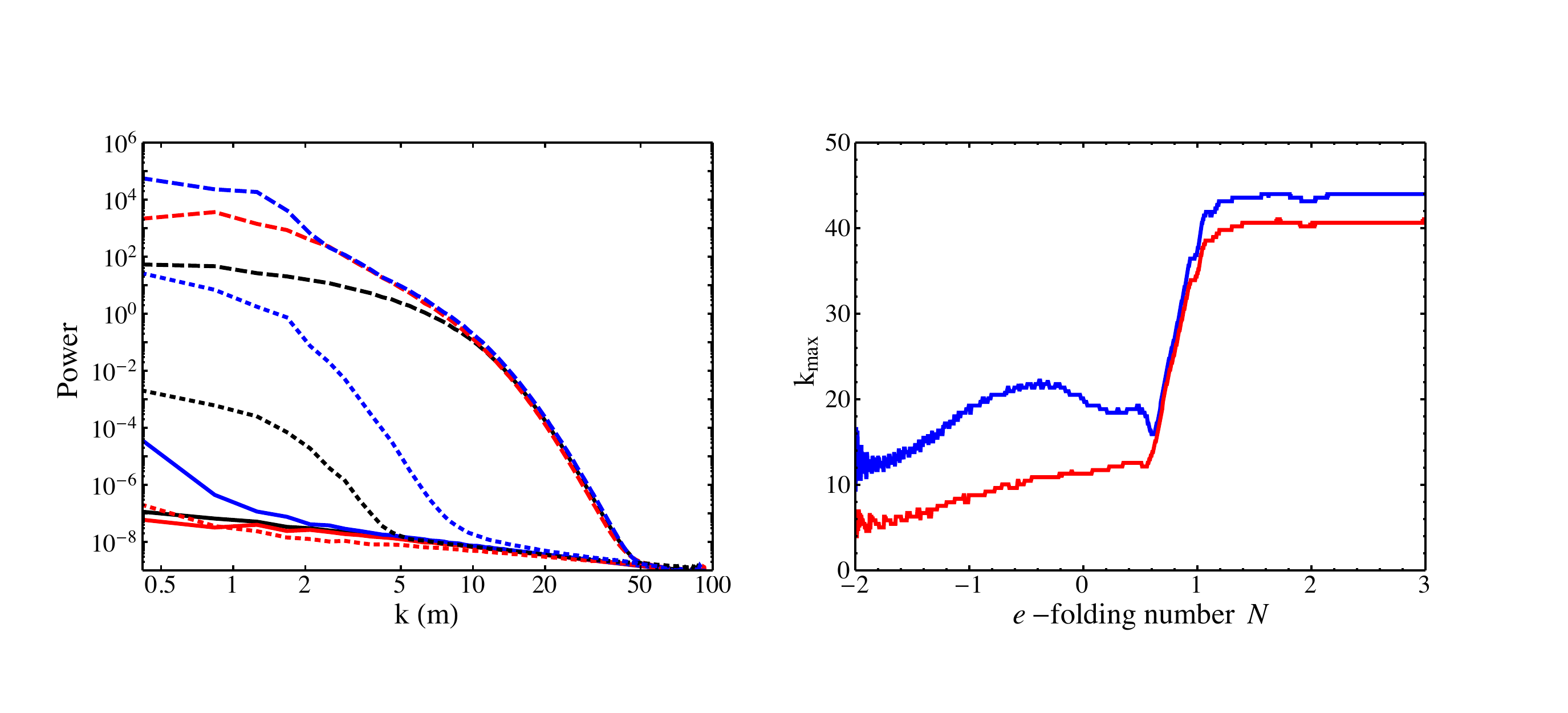}
\caption{The left panel shows spectra for $|A^-|^2$ (blue),  $|A^+|^2$ (red) and $\delta\phi^2$ (black)  for $N=-2$ (dashed) at $N=0$ (dotted) and $N=-2$ (solid). The right panel shows the maximum amplified wavenumber, defined as  $|A^-|^2> 4|A_{\rm BD}|^2$ (blue) and $|A^-|^2> 10|A_{\rm BD}|^2$ (red).}
\label{fig:kmax_60coupling}
\end{figure}

In the large coupling regime almost the entirety of the inflaton energy density is transferred to the gauge fields within one background oscillation, as shown in \cite{Adshead:2015pva}.\footnote{However, note that in this regime the axion field does not exhibit oscillations about the minimum of its potential, as described in the text.} As a characteristic coupling in this regime, we choose ${\alpha / f} = 60 \, m_{\rm Pl}^{-1}$. This is the region which is only accessible to lattice simulations (hence we do not perform a linear calculation), due to the significant backreaction and rescattering effects. The magnetic field strength obtained in this case --- see figure \ref{fig:parameterscan} --- is higher than that previously obtained in the literature \cite{Fujita:2015iga}, as expected due to the exponential dependence of the production of gauge fields on the coupling, as discussed in appendix \ref{app:gaugefieldsaxion}, and the fact that we were able to probe such high couplings using lattice simulations.  Due to the significant backreaction effects, the results of this region deserve further attention. In figure \ref{fig:kmax_60coupling} we plot the final gauge field spectra for the process in question. Based on the analysis of the tachyonic amplification, one would expect the maximum excited comoving wavenumber to be $k \simeq 12 \, m $. However, we see that in this case it is $k \simeq 50\, m $. This can be attributed to re-scattering of the gauge modes off the inflaton. It would present a problem to the predictive power of our results if the increase of $k_{\rm max}$ did not stop. In fact, we see the exact opposite, namely that $k_{\rm max}$ settles to a constant value after approximately $1.4$ e-folds after the end of inflation. The simulation lasts for more than $3.3$ e-folds after inflation ended, during which time $k_{\rm max}$ remains constant.

Before proceeding to further calculations, we briefly discuss the range of excited wavenumbers. Because the integral of the energy density would diverge if integrated over $k \in [0,\infty)$ due to the large volume of phase space in the ultraviolet, the spectrum must be renormalized to remove the unphysical effects of the short-wavelength Bunch-Davies contribution. We do this in a very simple and intuitive way. The range of integration is limited to include only wavenumbers that have had a significant amplification compared to their BD values. Specifically, we used two criteria: $|A^\pm|^2 > 10 |A_{\rm BD}|^2$ and $|A^\pm|^2 > 4 |A_{\rm BD}|^2$. As can be seen in figure \ref{fig:kmax_60coupling}, the lower cut-off gives a higher range of amplified wavenumbers $k_{\rm max}$ during inflation. As we saw in section \ref{sec:SmallCoupling}, our results are sensitive to the renormalization procedure at early times, before the end of inflation. At late times, since the low-$k$ modes are exponentially amplified, the difference in the definition of $k_{\rm max}$ does not affect the macroscopic quantities, such as $\xi_B$.

The large coupling case also exhibits a novel behavior regarding the helicity of the resulting gauge fields. While for low and moderate couplings the resulting gauge fields are strongly polarized at all scales, for large couplings we see two different regions in $k$-space. For large wavenumbers, rescattering is so efficient that the fields end in an almost unpolarized state. However,  a significant net polarization remains for low wavenumbers. This is important for the reheating behavior and post-reheating magnetic field evolution, as we explain in section \ref{sec:evolution}.
It is worth noting that if we artificially turn off the backreaction of the gauge fields onto the inflaton, we regain the expected range of excited wavenumbers and the dominant polarization is several orders of magnitude larger than the other one, as one would have calculated using analytic or semi-analytic (e.g. WKB) methods. The comparison of the full simulation to one with artificially suppressed backreaction was presented in \cite{Adshead:2015pva}.

Finally,  figure \ref{fig:parameterscan} shows an exponential dependence of the intensity of the produced hyper-magnetic field for small couplings and a common asymptotic behavior for large couplings. This is expected, since for large couplings almost the totality of the axion energy density is transferred to the gauge fields, hence for even larger couplings no further amplification of the hyper-magnetic field is possible. We can estimate the absolute theoretical maximum magnetic field that is possible, through equating the energy density at the end of inflation  $\rho_{\rm axion}  = H^2  \,3 m_{\rm Pl}^{2} / 8\pi \approx  0.04 \,m^2    m_{\rm Pl} ^2$ to the hyper-electromagnetic energy density $\rho_{EM} = 2B^2$, assuming instantaneous energy transfer between the axion and hypercharge fields, leading to
\begin{align}
{B_{\rm max}^2 \over m^4} \sim \left( 0.2 { m_{\rm Pl} \over m}   \right )^2  = {\cal O} \left (  10^{10}\right ) .
\label{eq:Bmaxmax}
\end{align}
Figure \ref{fig:parameterscan} shows that this is within an order of magnitude of the actual maximum value of the magnetic field produced in the large coupling regime, which is all we could anticipate by using this crude approximation.

\begin{figure}[t]
%\centering
\includegraphics[width=\textwidth]{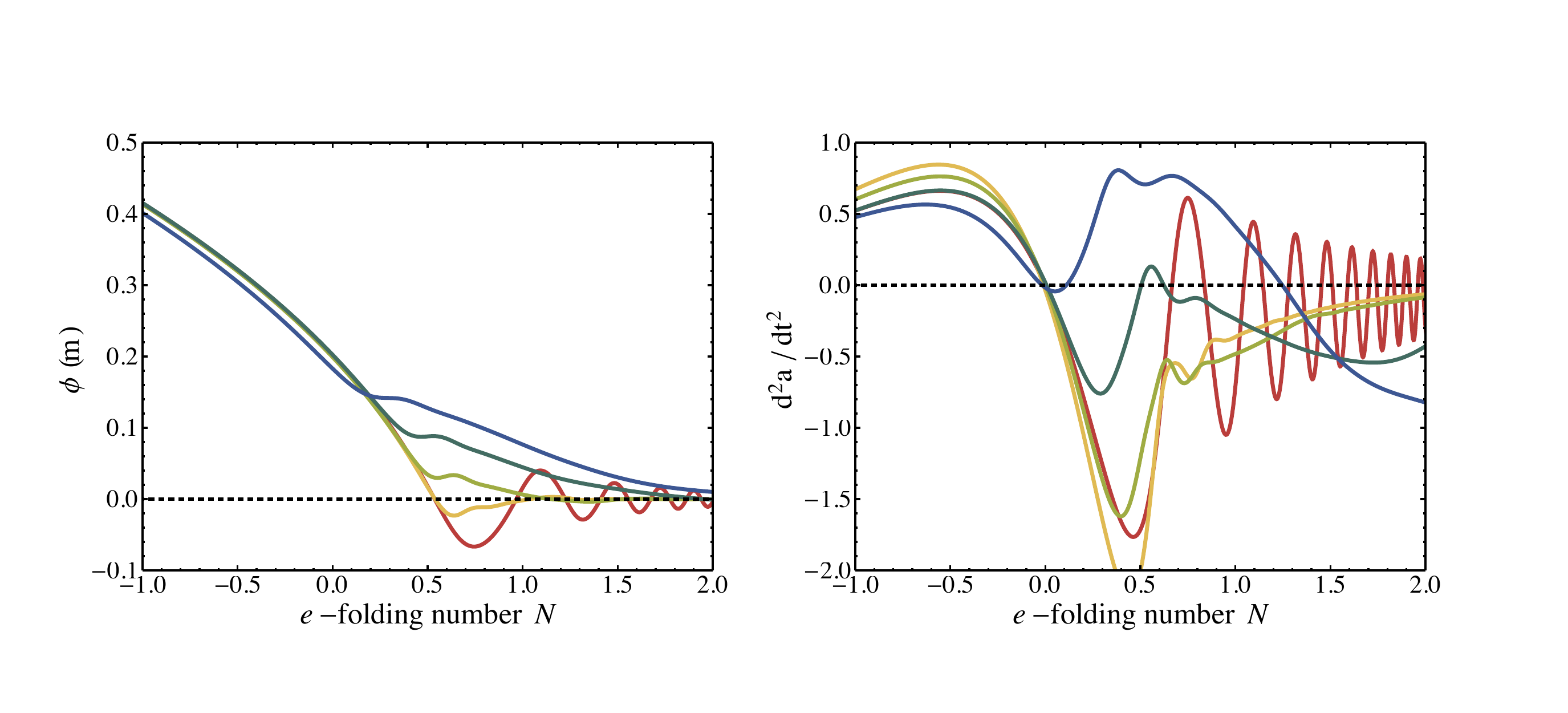}
\caption{The left panel shows average axion field value for $\alpha /f = 35\,m_{\rm Pl}^{-1}$, $\alpha /f =55\,m_{\rm Pl}^{-1}$, $\alpha / f=60\,m_{\rm Pl}^{-1}$, $\alpha /f =65 \,m_{\rm Pl}^{-1}$, and $\alpha / f=70 \,m_{\rm Pl}^{-1} $ (from red to blue color-coded on a rainbow scale). The black-dotted line corresponds to $\phi =0$. 
The right panel shows the acceleration of the expansion $\ddot a(t)$ for the same couplings. The black-dotted line corresponds to $\ddot a =0$. }
\label{fig:plateau}
\end{figure}

A natural question to ask is whether further increasing the coupling beyond $\alpha / f = 60\, m_{\rm Pl}^{-1}$ will bring the produced hypermagnetic field even closer to the theoretical maximum of eq.\ \eqref{eq:Bmaxmax}. The answer is more complicated than expected due to the non-linear nature of the system and the existence of strong backreaction effects that manifest earlier and earlier as the coupling is increased. Figure \ref{fig:plateau} shows the spatially averaged inflaton field value $\langle \phi (\vec x,t) \rangle_{\vec x}$, which is the lattice equivalent of the classical background field $\phi(t)$. As the gauge field coupling is increased, we see that the inflaton undergoes a brief period of ``trapping'' due to the non-linear interactions with the gauge fields. This trapping causes the inflaton to oscillate about a point on its potential away from the minima before slowly relaxing to its minimum. For $\alpha / f \lesssim 65 \,m_{\rm Pl}^{-1}$, the backreaction of the gauge fields is such that this trapping occurs after inflation has ended. However, for  $\alpha / f = 70 \,m_{\rm Pl}^{-1}$ these effects manifest \emph{during} inflation. This trapping stops inflation momentarily. During this interval, the inflaton field stops rolling and the gauge fields dominate the energy-density. The gauge fields then redshift, releasing the axion, which continues continues to roll and restarts inflation. In this example, the Universe inflates for about one more $e$-fold before inflation ultimately ends.
 In this case the first time that $\ddot a=0$ is not the end of inflation. This resembles models including sharp potential features, where inflation might momentarily stop, only to continue after the inflaton field has transversed the potential feature. We note that the convention of figure \ref{fig:plateau} denotes by $N=0$ the moment when $\ddot a=0$ for the first time. 

The existence of a plateau in the evolution of the background axion $\phi(t)$ indicates that the velocity is vanishing or becoming very small $\dot \phi \approx 0$ in this region (see figure \ref{fig:plateau}), and is indicative of extremely strong backreaction. This strong backreaction \emph{during inflation} will lead to an enhancement of the density perturbations that exit the horizon at that time, which for $\alpha / f \simeq 70 \,m_{\rm Pl}^{-1}$ is about one e-fold before the end of inflation.  Naively, in the time-delay formalism  \cite{Hawking:1982cz,Guth:1982ec,Guth:2013epa}, where $\delta \tau = {\delta \phi / \dot \phi}$ is the time-delay field, which translates to density perturbations $\delta \rho /\rho$, a point of vanishing inflaton velocity appears to be singular, and suggests large density perturbations.  However, in models where more than one degree-of-freedom is active during inflation, perturbations are more usefully defined through $\zeta$, which in spatially flat gauge is given by $\zeta =  \delta \rho /3 (\bar\rho +\bar p)$. On the one hand, the energy density at the moment when the axion momentarily stops rolling will be have a large contribution from the gauge fields which will prevent the background  $(\bar\rho +\bar p)$ from vanishing. On the other hand, the presence of the strong backreaction is likely to generate very large fluctuations in the density, $\delta\rho$, which may lead to the formation of primordial black holes \cite{Linde:2012bt, Bugaev:2013fya}. 
A detailed study of black hole production in this model is beyond the scope of this work. We thus restrict the large coupling regime for the remainder of this work to $\alpha / f < 70 \,m_{\rm Pl}^{-1}$, where there is no nonlinear axion trapping during inflation. 

%%%%%%%%%%%%%%%%%%%%%%%%%%%%%%%%%%%%%%%%%%%%%%%%%%%
%%%%%%%%%%%%%%%%%%%%%%%%%%%%%%%%%%%%%%%%%%%%%%%%%%%

\section{Post-reheating magnetic field evolution}
\label{sec:evolution}

In this section we begin by sketching the evolution of magnetic fields in the turbulent regime in magneto-hydro-dynamics (MHD). We present a calculation for the evolution of helical magnetic fields in a turbulent plasma in an idealized scenario. We then compare the spectra of the magnetic energy and helicity densities that arise in our model and argue for the validity of using the results of helical evolution found in the literature for our case. Finally, we estimate the effect of the reheat temperature on the present-day magnetic field.

\subsection{Helicity-dependent MHD evolution}

As demonstrated in section \ref{sec:Gaugefieldproduction}, magnetic fields in free space simply decay as $B_i \sim a^{-2}$  due to the expansion of space-time. This is the standard behavior of radiation modes in an expanding Universe. However, after reheating has ended, the Universe is filled with a plasma of charged particles. In this case, the evolution can be very different. The details depend on the properties of the plasma, such as its conductivity and magnetic Reynolds number. In this work, we are interested in the evolution of magnetic fields in the primordial plasma in its turbulent era, after reheating and before recombination. Numerical studies \cite{Banerjee:2004df} have explored the behavior of magnetic fields in different states of primordial plasma. Our analysis employs the results first presented in ref.\ \cite{Campanelli:2007tc}, where approximate analytical expressions for the role of helicity in the evolution of magnetic fields in the primordial plasma were given. As a means to outline the main logic and assumptions behind our analysis, we present the main steps of the calculation performed in ref.\ \cite{Campanelli:2007tc}, while directing the interested reader to ref.\ \cite{Campanelli:2007tc} for further details.

The equations for the magnetic energy and helicity spectra are governed by a system of ordinary differential equations
\begin{align}
\label{eq:helicityenergy1}
\partial_t {\cal E}_B  = &  -2 \eta_{\rm eff} k^2 {\cal E}_B  +\alpha_B k^2 {\cal H}_B
\\
\partial_t {\cal H}_B =  &  -2 \eta_{\rm eff} k^2 {\cal H}_B  +4 \alpha_B {\cal E}_B
\label{eq:helicityenergy2}
\end{align}
where the coefficients $\eta_{\rm eff}$ and $\alpha_B$ are given by
\begin{align}
\eta_{\rm eff} &= \eta  +4E_B {\tau_d \over 3},\quad \text{ and }\quad 
\alpha_B(t) = - \dot H_B  {\tau_d \over 3\eta}.
\end{align}
In these definitions $\tau_d$ is the fluid response time to the Lorentz force induced by the magnetic field and $\eta$ is the conductivity of the plasma.

To determine the evolution of a non-maximally helical field, we take the configuration ${\cal H}_B(k,0) = h_B 2k^{-1}{\cal E}_B(k,0)$ with $0<h_B<1$,  as an initial condition. This means that the helicity and energy spectra contain the same wavenumber information, apart from some overall magnitude.
In terms of the power spectrum this can be re-written as $P_A(k,0) = h_B P_S(k,0)$ leading to $|A_-(k,0)| \propto |A_+(k,0)|$ for all values of the wavenumber $k$.
The two helicity spectra have thus an identical $k$-dependence and differ only in their overall amplitude. In this case eqs.\ \eqref{eq:helicityenergy1} and \eqref{eq:helicityenergy2} can be formally solved for $  {\cal E}_B(k,t)$ and $  {\cal H}_B(k,t)$ as a function of their initial values, $ {\cal E}_B(k,0)$ and  ${\cal H}_B(k,0)$.
 For simplicity the initial spectrum was taken to be of the form ${\cal E}_B(k,0) = \lambda_B k^p \exp(-2k^2 l_B^2)$, thus following a power-law at small wavenumbers and falling off exponentially at large wave-numbers, so as to have finite energy.

By inserting the solutions for $ {\cal E}_B(k,t)$ and $ {\cal H}_B(k,t)$ into the definitions of eqs.\ \eqref{eq:EBdef} and \eqref{eq:HBdef}, the system of eqs.\  \eqref{eq:helicityenergy1} and  \eqref{eq:helicityenergy2} can be transformed into a system of equations for the integral quantities $E_B(t)$ and $H_B(t)$. There is significantly different behavior in the helical and the non-helical cases. In the non-helical case ($h_B=0$), the late-time evolution can be written as
\begin{align}
E_B(\tau) &\sim \tau^{-2(1+p)/(3+p)}
\\
\xi_B(\tau) &\sim \tau^{2/(3+p)}.
\end{align}
The product $E_B(\tau)\xi_B(\tau)$ thus decays as $\tau^{-2p/(3+p)}$. In the non-maximally helical case, one can distinguish two qualitatively different regimes. Initially, the system behaves as in the non-helical case, but once helicity starts to dominate energy is transferred from small to large scales through the inverse cascade. In the inverse-cascade region, the energy density and correlation length evolve as
\begin{align}
E_B(\tau) &\sim (\ln \tau)^{1/3} \tau^{-2/3}
\\
\xi_B(\tau) &\sim (\ln \tau)^{-1/3}  \tau^{2/3}.
\end{align}
The product is $E_B(\tau) \xi_B(\tau) = h_B E_B(0) \xi_B(0) = H_B(0)/2$. This saturates the integral condition of eq.\ \eqref{eq:realizInt}, and the system becomes maximally helical after entering the inverse cascade regime. By equating the product $E_B(\tau) \xi_B(\tau)$ in the two regimes, we can read off the time at which the helical evolution begins 
\begin{align}
t_h \simeq {1\over \sqrt {\kappa_{\rm diss}}} h_B^{-(3+p)/2p},\quad \text{where} \quad  \kappa_{\rm diss} = \gamma (1+p) \frac{\zeta_B^2(p)}{ 6},
\label{eq:helicaltime}
\end{align}
where  $\gamma$ is the asymptotic growth rate of the drag time and 
\begin{align}
\zeta_B(p) = \frac{\sqrt 2 \Gamma \left [p/2\right ]}{ \Gamma \left [ (1+p) /2 \right ]}, 
\end{align}
where $\Gamma[x]$ is the Euler Gamma function. An explicit simulation is shown in ref.\ \cite{Campanelli:2007tc} for the case of $h_B=10^{-3}$, where the analytic formulas were shown to describe the evolution of the system well.
\begin{figure}[t]
\includegraphics[width=\textwidth]{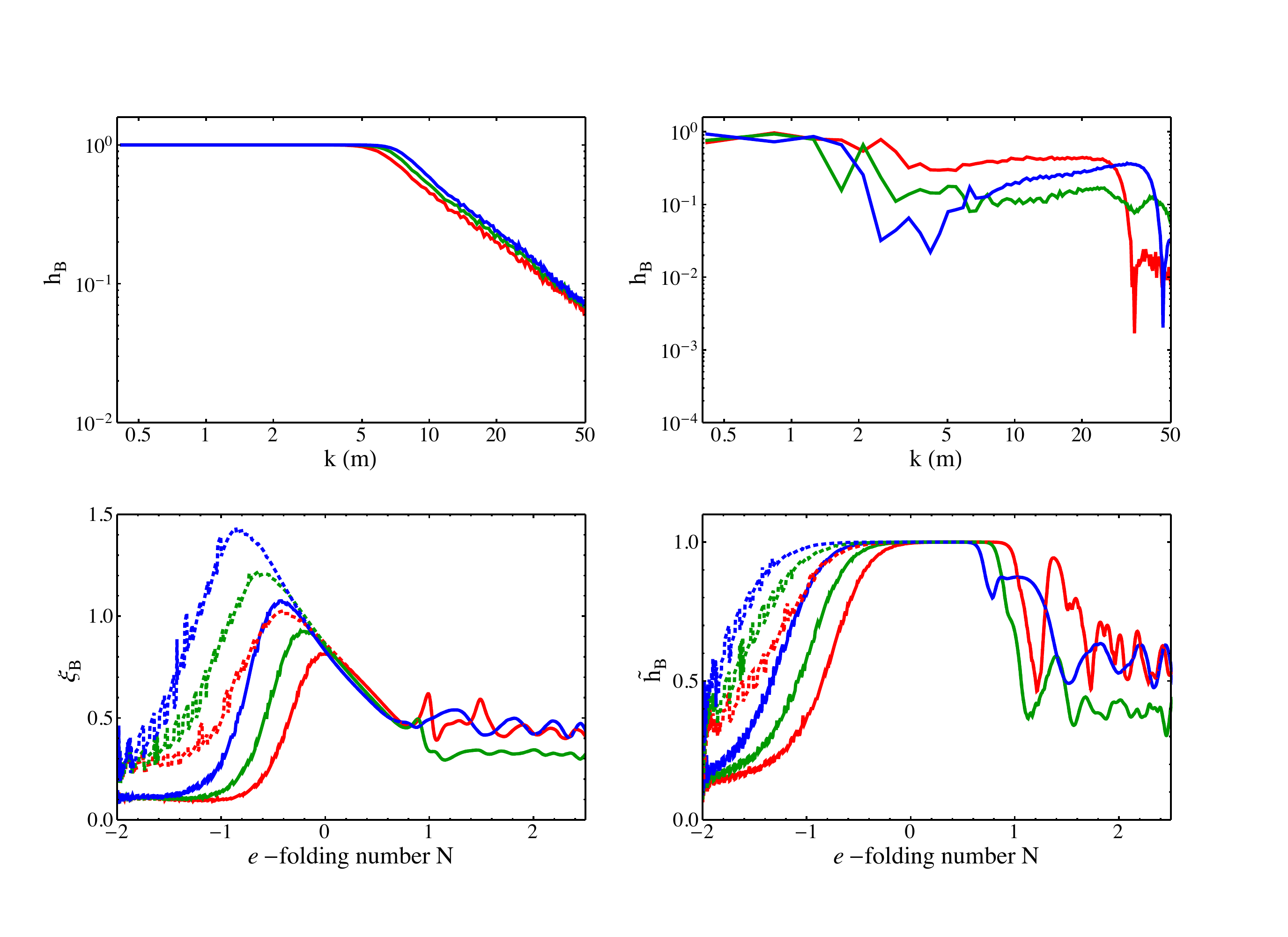}
\caption{The helicity fraction $h_B =  {\cal H}_B / (2 k^{-1}{ \cal E}_B)$ at the end of inflation  (upper left). The helicity fraction $h_B =  {\cal H}_B / (2 k^{-1}{ \cal E}_B)$ at $N=2$ e-folds after the end of inflation (upper right). The comoving correlation length $\xi_B$ for $k_{\rm max}$ defined through $|A^\pm|^2>4|A_{\rm BD}|^2$ (solid) and $|A^\pm|^2>10|A_{\rm BD}|^2$ (dotted) as a function of e-folding number (lower left).  The integrated helicity fraction $\tilde h_B = H_B / (2\xi_B E_B)$ as a function of e-folding number (lower right) with solid and dotted curves corresponding to $|A^\pm|^2>4|A_{\rm BD}|^2$ and $|A^\pm|^2>10|A_{\rm BD}|^2$ as before. 
In all panels the couplings are $\alpha / f  = 50\,m_{\rm Pl}^{-1}$ (red), $\alpha/f = 55\,m_{\rm Pl}^{-1}$ (green) and $\alpha/f = 60\,m_{\rm Pl}^{-1}$ (blue).
}
\label{fig:helicity_60coupling}
\end{figure}
For large couplings, as shown in ref.\ \cite{Adshead:2015pva}, the power spectra of the two helicity modes coincide for large $k$ and diverge for low $k$. This means that the energy and helicity density spectra are not related by a constant $h_B$, as in the analysis presented above. However, we use the results of this analysis, where we define the degree of partial helicity $\tilde h_B$ through the integral equation $H_B = \tilde h_B \times 2 \xi_B E_B$.

Let us now take a closer look at the helicity of the produced hypermagnetic fields in the case of axion inflation. We  focus on the large coupling case, since for low couplings the produced spectra are predominately helical. We use the results for simulations with couplings $(\alpha/f) m_{\rm Pl} \in [50,60] $ as the region of interest. Figure \ref{fig:helicity_60coupling} shows all relevant quantities. We see that the helicity parameter
$h_B = {\cal H}_B / {2 k E_B}$ is equal to unity for all relevant wavenumbers at the end of inflation. This is before re-scattering effects have allowed the generation of photons with the opposite helicity. At late times ($N=2$)  we see that $h_B \approx 1 $ for $k \lesssim 2 m$ and then it drops by different amounts for different values of the coupling. For ${\alpha / f} = 50 \, m_{\rm Pl}^{-1}$ and ${\alpha / f} = 55\,m_{\rm Pl}^{-1}$ the spectra are polarized at all scales, having $ h_B>0.1$. Further increasing the coupling pushes the helicity fraction $ h_B$ down to the $1\%$ level for large wavenumbers.
 
The correlation length shows the range of wave-numbers where most of the energy density is concentrated. First of all, we see that regardless of the exact definition of the renormalization criterion, the correlation length is the same after the end of inflation. Furthermore, it is very similar for all large-coupling runs and settles around a value of $\xi_B \approx 0.5\, m^{-1}$, corresponding to a  wavenumber of $k\approx 2m$. This means that the important part of the spectrum is significantly polarized. The integrated helicity ratio $\tilde h_B  = H_B  / (2 \xi_B E_B) $ settles to a value $\tilde h_B \sim 0.5$ at the end of preheating for the large-coupling runs with $50m_{\rm Pl}^{-1} \le   \alpha/f \le 60 m_{\rm Pl}^{-1}$, as shown in figure  \ref{fig:helicity_60coupling}.

Using eq.~\eqref{eq:helicaltime}, we can see that the helical evolution of the system starts at $t_h = {\cal O}(1) / \sqrt{\kappa_{\rm diss}}$. We ignore this small time offset and consider the helical evolution to start immediately. It would be interesting to perform an MHD calculation using a magnetic field spectrum of the present form and calculate how the helical and non-helical parts of the spectrum evolve in a charged plasma. We do not, however, expect the results to alter our qualitative understanding of the processes involved.

%%%%%%%%%%%%%%%%%%%%%%%%%%%%%%%%%%%%%%%
%%%%%%%%%%%%%%%%%%%%%%%%%%%%%%%%%%%%%%%
%%%%%%%%%%%%%%%%%%%%%%%%%%%%%%%%%%%%%%%

\subsection{Reheat temperature}

For a given inflationary scale, a faster transfer of energy from the inflaton into radiative degrees of freedom results in a higher reheat temperature. Thus the reheat temperature is an indicator of the speed of the transition between inflation and a radiation dominated Universe. 
Ref.\ \cite{Fujita:2015iga} showed that the late-time magnetic field produced in axion inflation coupled to a $U(1)$ field
increases with the square root of the reheat temperature, for couplings $\alpha / f\lesssim 40 m_{\rm Pl}^{-1}$, where instantaneous preheating does not occur. In that context a radiation-dominated Universe coincided with a Universe filled with a charged plasma.
 Hence for a fixed inflationary energy scale, a higher reheat temperature indicates that the Universe is filled with a charged plasma sooner following the end of inflation. This allows the inverse cascade process to start earlier and  the magnetic fields have less time to red-shift away after the end of inflation. Therefore, in order to calculate the present-day amplitude of the magnetic fields we first need to examine how quickly the Universe becomes radiation-dominated and (subsequently) filled with a charged plasma. 

\subsubsection{Low Coupling}
For low coupling values $\alpha/ f \lesssim 40 m_{\rm Pl}^{-1}$, tachyonic and parametric resonance is not strong enough to transfer the entirety of the inflaton energy density to gauge fields $A^\mu$ \cite{Adshead:2015pva}. In this case the inflaton can decay to gauge fields perturbatively. The relevant decay rate is
\begin{align}
\Gamma_{\phi\to AA} = {\alpha^2 m^3 \over 64 \pi f^2}.
\end{align}
The reheating temperature is obtained by demanding the decay rate equal to the expansion rate $\Gamma / 3H \sim 1$
\begin{align}
T_{\rm reh}\sim 1.31 \times 10^8 \left ( {100\over g_*} \right )^{1/4}   \left ( { \alpha \over  f m_{\rm Pl}} \right ) \left  ({m \over 1.06 \times 10^{-6} m_{\rm Pl}} \right )^{3/2} {\rm GeV} \, ,
\end{align}
where we considered chaotic inflation with a  quadratic potential and $g_*$ is the effective number of relativistic degrees of freedom.

The shift symmetry of the axion severely restricts the coupling to other fields. The two allowed dimension-5 couplings are the axion-gauge coupling analyzed here and the axion-fermion coupling 
\begin{align}
{\cal L}_{\rm int} = i {C\over f} \partial_\mu \phi \,\bar \psi \gamma_5 \gamma^\mu \psi
\end{align}
where $\gamma^\mu  = e^\mu_a \gamma^a $ and $e^\mu_a$ is the vielbein, which  encoded the effects of the curved space-time. Despite the fact that this coupling can lead to non-perturbative production of fermions during and after inflation \cite{Adshead:2015kza, Adshead:2015jza}, the Pauli exclusion principle does not allow for this process to transfer a significant amount of the inflaton's energy to the fermions. Hence, we must again consider perturbative reheating. The decay rate is
\begin{align}
\Gamma_{\phi \to \psi \bar \psi} \propto  \( { C\over f} \)^2 m_\psi^2 \, m \, ,
\end{align}
where $m_\psi$ is the fermion mass.  Because of the derivative coupling, the decay rate is suppressed by $(m_\psi / m)^2$, the reheating temperature is lowered by $m_\psi / m$, since $T_{\rm reh} \sim \sqrt{ \Gamma M_{\rm Pl}}$.

\subsubsection{Large Coupling}

Calculating the reheating temperature in the large coupling regime is somewhat more interesting.  Here, the entirety of the inflaton energy is transferred to the gauge fields much faster than the usual perturbative decay channels into gauge bosons or fermions can operate, due to the effectiveness of tachyonic preheating. Here, shortly after inflation, the Universe ends up filled with radiation in the form of $U(1)$ hypercharge bosons, with a helicity asymmetry. This is far from a primordial plasma and some decay channel to charged particles must be invoked in order to connect this picture to the standard hot big bang. 

In order to study the interactions between the hypercharge sector and charged fermions, we start by considering the full electroweak Lagrangian, which contains
\begin{align}
{\cal L}_{\rm EW} \supset   \left | D_\mu \Phi \right |^2 =  \left |  \left (\partial_\mu - i g W^a_\mu \tau^a - i {1\over 2} g' A_\mu \right ) \Phi \right |^2,
\label{eq:electroweak}
\end{align}
where $\Phi$ is the Standard Model Higgs.

\begin{figure}[t]
\centering
\includegraphics[height = 1.5in]{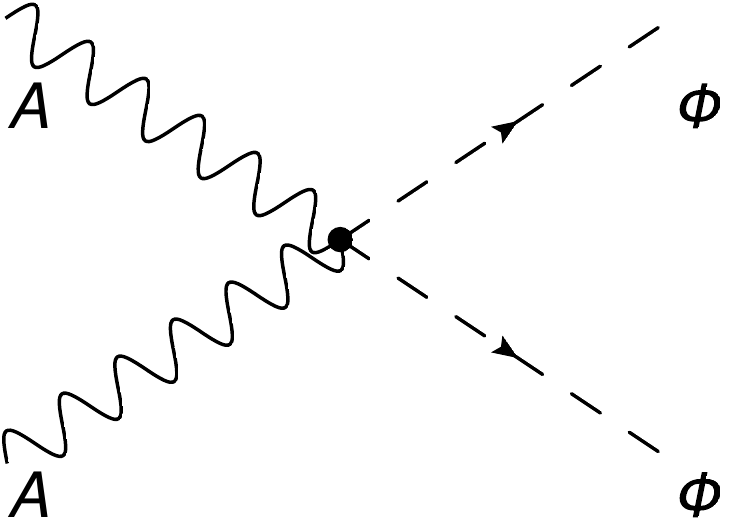}  \hspace{1in} 
\includegraphics[height = 1.5in]{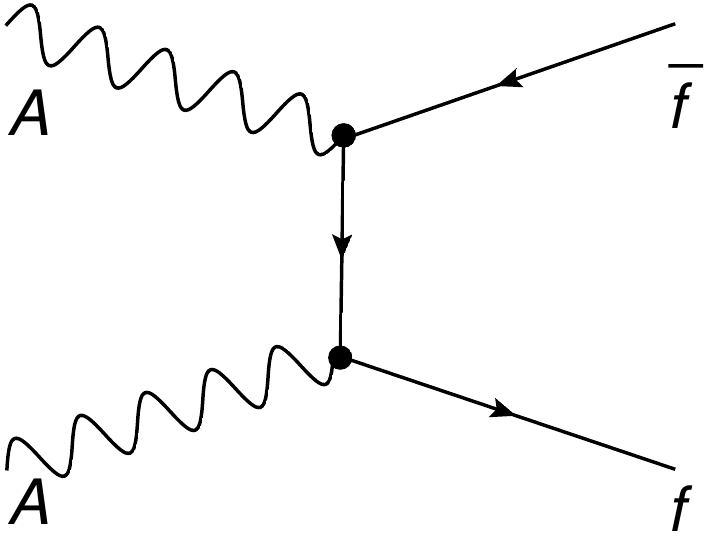}
\caption{  Scattering vertices that allow the transfer of energy from the hypercharge bosons to the rest of the Standard Model. }
\label{fig:BB_PhiPhi}
\end{figure}

We can write the Higgs field in its general form as
\begin{align}
\Phi =   \( \begin{array}{c}
\varphi  \\ \varphi_0  
\end{array} \) 
\end{align}
where $\varphi = \varphi_1 + i \varphi_2$ and $\varphi_0 = \varphi_3 + i \varphi_4$ are complex functions.\footnote{We use $\phi$ for the axion-inflaton and $\varphi$ for the components of the Higgs field.} After the electroweak phase transition we write $\varphi =0$ and $\varphi_0 = (v+ h)/\sqrt 2$ (with $v$ and $h$  real), which gives the known interactions of $Z$ and $W^\pm$ bosons to the remaining real part of the Higgs. 
For our purpose, we expand eq.\ \eqref{eq:electroweak} as
\begin{align}
\nonumber
{\cal L}_{\rm EW} \supset &   {g g'\over 4} \left  [ W^1_\mu A^\mu (\varphi^* \varphi_0  + \varphi_0^* \varphi) + W^2_\mu A^\mu ( -i \varphi^* \varphi_0 + i \varphi \varphi_0^*) + W^3_\mu A^\mu (|\varphi|^2 - |\varphi_0|^2)  \right ]
\\
& { (g')^2\over 8} A_\mu A^\mu (|\varphi|^2  + |\varphi_0|^2) + {g^2 \over 8}  W_\mu^a W^{b\mu}  (\varphi^* \varphi_0^*)\sigma^a \sigma^b \( \begin{array}{c}\varphi  \\ \varphi_0  \end{array} \) .
\label{eq:ewL}
\end{align}
We only need to estimate the rate of the process $A A \to \Phi \Phi $, shown in the left panel of figure  \ref{fig:BB_PhiPhi}. This process is facilitated by the interaction ${\cal L}_{\rm int} = [(g')^2 /8 ] A_\mu A^\mu (|\varphi|^2  + |\varphi_0|^2) $, which arises, for example, in the context of scalar QED.
This process has the matrix element
 \begin{align}
 i {\cal M} = i  {\alpha_Y}^2  g_{\mu\nu} \varepsilon_1^{*\mu}\varepsilon_2^{*\nu}
 \end{align}
 where $\varepsilon_1 , \varepsilon_2$ are the polarization vectors of the two incoming hypercharge bosons. Usually one would sum over all polarizations but in our case this is true only for the unpolarized part of the spectrum. For the helical part of the spectrum, we take both gauge bosons having the same circular polarization, which ends up giving the factor $g_{\mu\nu} \varepsilon_1^{*\mu}\varepsilon_2^{*\nu} \sim (1-\cos\theta)$, where $\theta$ is the angle between the two gauge bosons. Since we expect a spherically symmetric distribution, we must integrate over a full sphere in $k$ space, in which case the gauge bosons that are moving towards the same direction will have a suppressed cross-section, but the ones that move towards each other will not. This is intuitive, since two helical states must add up to a helicity zero  in order to produce two $s$-wave scalar particles. We do not pursue the calculation in detail since we are only interested in order-of-magnitude results for the scattering cross-section.
The rate of the $AA\to \Phi\Phi$ process is 
\begin{align}
\sigma_{AA\to \Phi\Phi } \sim {\alpha_Y^2 \over s}
\label{eq:sigmaBBphiphi}
\end{align}
since there are no other mass or energy scales in the problem apart from the center-of-mass energy of the incoming gauge bosons $s$. 
The relevant quantity that characterizes the efficiency of this process is the ratio of the scattering rate to the Hubble rate 
\begin{align}
{\Gamma \over H} = {n\sigma v\over H} \, .
\label{eq:GammaoverH}
\end{align}
We only perform order-of-magnitude calculations and try to clearly state our assumptions.
The particle density can be calculated as the ratio of the energy density $\rho \sim H^2 m_{\rm Pl}^2 \sim m^2 m_{\rm Pl}^2$ to the energy per particle, which is of the order of $H\sim m$. Altogether, $n \sim m\, m_{\rm Pl}^2$. 
Since $s \simeq m ^2 \simeq H^2$, eq.\ \eqref{eq:GammaoverH} becomes
\begin{align}
{\Gamma \over H} \sim \alpha_Y^2  \left ( {m_{\rm Pl} \over m} \right )^2 \gg 1.
\label{eq:gammaoverHresult}
\end{align}
The Higgs bosons $\varphi$ and $\varphi_0$ can further annihilate to produce charged fermions with a similar scattering rate (effectively instantaneously). In deriving eq.\ \eqref{eq:gammaoverHresult} we took the relevant particle momenta to scale as $k\sim m$. This is only correct for quadratic inflation. Considering a more general potential, since the maximum gauge field amplification occurs around the end of inflation, the relevant wavenumbers scale like $k \sim H_{\rm end}$, which for quadratic inflation is also proportional to the inflaton mass. This means that the end result is not sensitive to the specific form of the inflationary potential, as long as we use eq.\ \eqref{eq:gammaoverHresult} with the substitution $m \to H$.

The above calculation holds if the electroweak symmetry is unbroken during inflation, or broken and then thermally restored during reheating. In the opposite case, one can assume that a light Higgs during inflation acquires a VEV (or performs a random walk acquiring an root-mean-square value on the Hubble patch that is our Universe \cite{Enqvist:2013kaa}) and hence electroweak symmetry is broken with $v\sim H$. Here, the scattering cross section will be $\sigma \sim \alpha_W^2  / v^2 \sim \alpha_W ^2 / H^2 $, since the Higgs will have a mass and the incoming states will have a similar energy, putting the reaction $AA\to \Phi\Phi$ near resonance. We see that in both cases, broken or unbroken electroweak symmetry, the resulting cross-section is of the same order and much faster than the Hubble rate. 
The thermalization of the particles in the plasma will proceed via similar scattering events, hence we can also consider this to be instantaneous. 

The second vertex in figure \ref{fig:BB_PhiPhi} shows a direct channel from gauge bosons into charged fermions, without the need for the Higgs field to act as an intermediary. This scattering process is suppressed for a polarized initial state of gauge bosons and final states that are light with respect to the center-of-mass energy, and is forbidden if the resulting fermions are massless. However, the resulting spectrum of gauge bosons produced for large axion-gauge coupling values has a significant unpolarized component. Simple dimensional arguments give the cross-section of the process $AA \to f \bar f$ (where $f$ is some charged fermion) to be similar to the rate of eq.\ \eqref{eq:sigmaBBphiphi}, leading to the same instantaneous reheating effect.

Summing up the results and using the approximation of instantaneous reheating, the corresponding temperature is given by equating the energy density of the inflaton at the end of inflation to the energy density of a thermal gas of particles
\begin{align}
( 3m_{\rm Pl}^2 / 8\pi) H_{\rm end}^2  = \rho_{\rm rad} = \sigma_{\rm SB} T_{\rm reh} ^4
\end{align}
where $\sigma_{\rm SB} = \pi^2 / 60$ is the Stefan-Boltzman constant, giving the reheat temperature
\begin{align}
\label{eq:Treheat}
T_{\rm reh} =\left ( {3\over \sigma_{\rm SB} }  \right )^{1/4} \sqrt {{m_{\rm Pl}\over \sqrt{8\pi}} H_{\rm end} }  \sim \sqrt {     m \times m_{\rm Pl}} \,. 
\end{align}

\subsection{Late-Universe Magnetic Field}

We now have all the tools needed to estimate the intensity and correlation length of the produced magnetic field as it would be measured in the late Universe. We do not know what fraction of the hypercharge fields scatter into charged particles. We can parametrize the unknown fraction of the remaining energy density in $U(1)_Y$ fields by a parameter $\varepsilon_B < 1$. The transformation of the hypermagnetic to magnetic field occuring at the electroweak transition has a high efficiency $(\cos \theta_w \sim .9)$ \cite{Dimopoulos:2001wx}. We include this parameter in $\varepsilon_B $.

For simplicity we consider that only the magnitude of the magnetic field spectrum is suppressed by $\varepsilon_B$, while its shape, and its correlation length, remains unaffected. It was shown in ref.\ \cite{Fujita:2015iga} that the intensity of the magnetic field produced for ${\alpha / f} \lesssim 40m_{\rm Pl}^{-1}$ is not sufficient to explain the blazar observations, we thus concentrate on the case ${\alpha / f} > 40m_{\rm Pl}^{-1}$, where tachyonic preheating leads to a complete transfer of the energy from the inflaton to the hypercharge fields \cite{Adshead:2015pva} and lattice simulations are unavoidable, due to the large backreaction.

We  perform the calculation for the case $\alpha / f = 60 m_{\rm Pl}^{-1}$ in detail and all other large-coupling cases can be derived from that, using the results of figure \ref{fig:parameterscan}. The key point of the large-coupling regime is instantaneous reheating with a temperature given in eq.\ \eqref{eq:Treheat}. We take the conductivity of the primordial plasma after reheating to scale as $\sigma \simeq 100 T$ \cite{Arnold:2000dr, Baym:1997gq}. The magnetic Reynolds number is defined $R_m(k) = v_k \sigma /k_{\rm phys}$ for each physical wave-number $k_{\rm phys}$. For instantaneous reheating, the maximum relevant wavenumber is 
\begin{align}
k_{\rm phys} = {\cal O}(1) {\alpha\over f} \left ( m_{\rm Pl}\right )  H_{\rm end}\,  a(t) ,
\end{align}
where $H_{\rm end}$ is the Hubble scale at the end of inflation, and $a(t)$ is the scale-factor normalized so that $a(t)=1$ at the end of inflation. The scale-factor in a radiation dominated Universe is related to temperature as $a(t)  \propto T_{reh} / T$. The proportionality factor is an ${\cal O}(1)$ number in the case of instantaneous reheating, taking the form $e^N$, where $N$ is the number of e-folds between the end of inflation and the transition of the Universe to a state filled with a charged plasma.
Putting everything together into the definition of the  magnetic Reynolds number, we have, 
\begin{align}
R_m = {\cal O}(1)  \sqrt{
m_{\rm Pl}
\over 
H_{\rm end} 
} \, v_k,
\end{align}
where we took $50 \lesssim (\alpha/f) m_{\rm Pl}  \lesssim 60$ and combined all numerical factors into the ${\cal O}(1)$ pre-factor. To account for the fact that $H_{\rm end} / m_{\rm Pl}\simeq 10^{-6}$, we need to take fluid velocity $v_k>10^{-3}$ to have $R_m>1$. We assume that this condition holds for most of the period of interest, so that the magnetic fields evolve in a turbulent plasma.

We then use the fact that the helicity of the magnetic field is conserved between the end of reheating and the present day
\begin{align}
a^3(t_{\rm reh}) \lambda_{\rm phys}(t_{\rm reh}) B_{\rm phys}^2(t_{\rm reh}) = a^3(t_{\rm late}) \lambda_{\rm phys}(t_{\rm late}) B_{\rm phys}^2(t_{\rm late}),
\end{align}
where the subscript ``reh'' refers to the physical quantities after the hypercharge bosons have scattered into charged particles and the subscript ``late'' refers to the late Universe, where we are interested in the magnetic field as a seed for the galactic dynamo or an explanation for the blazar observations. 

We solve for the combined quantity 
\begin{align}
B_{\rm eff} = B_{\rm phys}(t_{\rm late}) \sqrt { \lambda_{\rm phys}(t_{\rm late}) \over 1\, {\rm Mpc}},
\end{align}
Altogether this leads to 
\begin{align}
B_{\rm eff} = \left ({ a_{\rm reh} \over a_{\rm late} }\right )^{3/2}\sqrt { \lambda_{\rm phys}(t_{\rm reh}) \over 1\,{\rm Mpc}} B_{\rm phys}(t_{\rm reh}).
\end{align}
\begin{figure}[t]
\centering
\includegraphics[width=\textwidth]{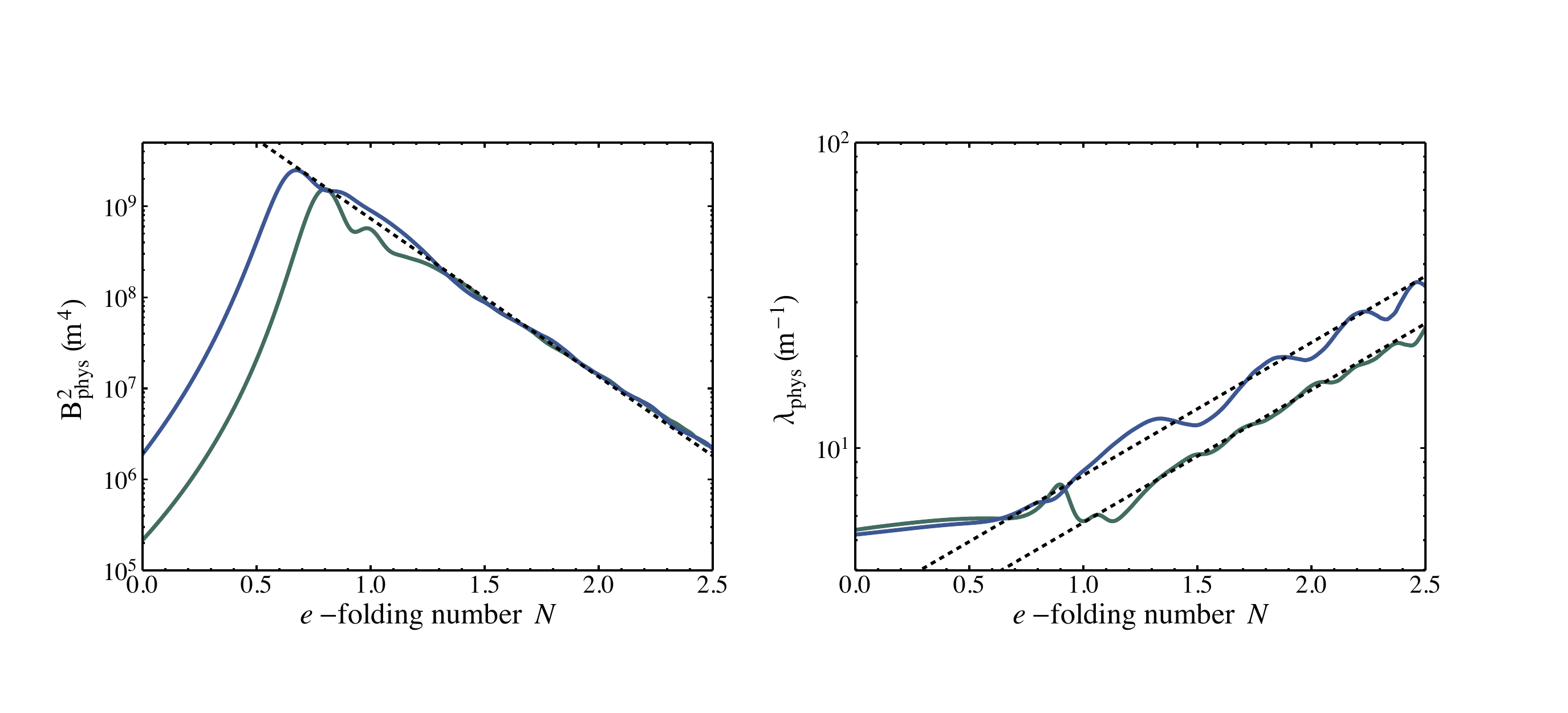}
\caption{The physical magnetic field and correlation length for $\alpha=55m_{\rm Pl}^{-1}$ (dark green) and $\alpha=60m_{\rm Pl}^{-1}$ (blue) along with the late-time exponential fitting curves.  }
\label{fig:magfield60}
\end{figure}
We now fit the late-time behavior of $B^2$ and $\lambda$ by the functions  
\begin{align}
\lambda_{\rm phys}(t_{\rm reh} )&=3.3e^N m^{-1}=20 \cdot 10^{-52} e^N \, {\rm Mpc}
\\
B^2 _{\rm phys}(t_{\rm reh} ) &= 5.5 \cdot 10^{10} e^{-4N} m^4= 3.3 \cdot 10^{101} e^{-4N} \, {\rm G}^2 \, ,
\end{align} 
as shown in figure \ref{fig:magfield60}. Since in the large coupling regime these fitting functions do not strongly depend on the coupling, we use the numerical fit factors for $\alpha / f = 60 m_{\rm Pl}^{-1}$. The results do not differ significantly for other large couplings.
By using these fitting functions and inserting the expansion of the Universe from the end of reheating until today, which is $\sim 10^{26}$, the effective magnetic field becomes
\begin{align}
B_{\rm eff} \simeq 2.5 \cdot   10^{-14} \cdot e^{-3N_{\rm reh}/2} \, {\rm G}.
\end{align}
The number $N_{\rm reh}$ is the number of e-folds  where the redshifting of the magnetic fields starts. From figure \ref{fig:magfield60}, we can estimate $N\simeq 1$, making the exponential term about $0.2$. There is one further suppression, $\varepsilon_B \leq {\cal O}(1)$, so that $B_{\rm eff} \gtrsim 10^{-16} $. The lower bound for explaining the blazar observations is either $B_{\rm eff}\ge 10^{-15}$ or $B_{\rm eff}\ge 10^{-17}$ depending on  assumptions  \cite{Taylor:2011bn}. This makes axion inflation a serious possibility for explaining the observed cosmological magnetic fields.

Although $B_{\rm eff}$ is enough to compare the model's predictions with observations, since we  expect $\lambda_{\rm phys}$ to be less than $1$ Mpc, it is worth trying to disentangle the two quantities, the magnetic field and the correlation length, and calculate their present-day value. We make the standard assumption that turbulent evolution and the inverse cascade continues until recombination, where the Universe becomes largely neutral. After that, the magnetic field and correlation length simply redshift as $B_{\rm phys}^2 \sim a^{-4}$ and $\lambda_{\rm phys} \sim a$. During the turbulent evolution, equipartition is achieved between the plasma kinetic energy and the magnetic field. The correlation length scales as $\lambda \sim v_A t$ where $t$ is cosmic time and $v_A\sim B / \sqrt{\rho}$ is the Alfven speed. This gives immediately
\begin{align}
B_{\rm rec} \sim \lambda_{\rm rec} { \sqrt{\rho_{\rm rec}} \over t_{\rm rec}} ,
\end{align}
and the evolution from recombination until today is simply performed through by multiplying with a factor of $(a_{\rm rec} / a_{\rm present})^3$. 

Careful analytic and numerical calculations refining the train of thought above (see refs.\ \cite{Banerjee:2004df, Durrer:2013pga} and references therein) give the relation of the present-day magnetic field and correlation length as
\begin{align}
B_{\rm phys}(t_{\rm late}) \sim 10^{-8} \left (   { \lambda_{\rm phys}(t_{\rm late}) \over 1\, {\rm Mpc}  }  \right ) {\rm G}.
\label{eq:Blambda_dec}
\end{align}
Using eq.\ \eqref{eq:Blambda_dec}, along with the value of $ B_{\rm eff} \gtrsim 10^{-16} \,{\rm G}$, we can calculate
\begin{align}
B_{\rm phys}(t_{\rm late}) \sim 10^{-13}\, {\rm G},   \quad \quad \lambda_{\rm phys}(t_{\rm late}) \sim 10\, {\rm pc} \, .
\end{align}
The current physical correlation length is clearly smaller than the galactic scale but the strength of the magnetic field amplitude leads to a $B_{\rm eff}$ that can be relevant for blazar observations. 

Recently a connection between primordial magnetic fields and baryogenesis was proposed through the non-conservation of magnetic helicity, due to the finite conductivity of the primordial plasma \cite{Anber:2015yca,Fujita:2016igl}. It is shown in \cite{Fujita:2016igl} that the intensity of the late-time magnetic field needs to lie in the interval $  10^{-14} {\rm G} < B_{\rm phys} < 10^{-12} {\rm G}$ to produce the observed baryon asymmetry.
\footnote{Shortly after the completion of the present work, it was shown in \cite{Kamada:2016eeb} that the resulting baryon asymmetry for strong magnetic field (as the one generated in this model) is suppressed compared to previously calculated values, due to the chiral magnetic effect. This indicates that is is harder --if at all possible-- to generate both sufficiently strong intergalactic magnetic fields and the observable baryon asymmetry using primordial magnetic fields in the context of axion inflation.}
 The predicted value  of $B_{\rm phys}(t_{\rm late}) \sim 10^{-13}\, {\rm G}$ predicted by our calculations is at the center of this range. Hence the coupling of the $U(1)_Y$ field to axion inflation can be a viable solution to both magnetogenesis and baryogenesis, provided that the coupling strength is large enough to allow for instantaneous reheating, $\alpha / f \simeq 60 m_{\rm Pl}^{-1}$.

\section{Conclusions}
\label{sec:conclusions}

In this paper we have studied the  production of hyper-magnetic fields following a period of axion-driven inflation in the early Universe. In this scenario, the axion-inflaton is coupled to $U(1)$ gauge fields via a dimension-$5$ interaction of the form ${\cal L}_{\rm int} \sim \phi F_{\mu\nu}\tilde F^{\mu\nu}$. As is well-known, this leads to the production of gauge fields starting during inflation and continuing through preheating. We identify the abelian gauge field as the $U(1)_Y$ hypercharge field of the Standard Model and calculate the resulting large-scale hyper-magnetic field.

Using the methods described in \cite{Adshead:2015pva}, we use lattice simulations to self-consistently calculate the production of hypermagnetic fields at the end of axion inflation.   Immediately following the end of inflation, at large enough values of the coupling, near-instantaneous preheating leads to a radiation-dominated Universe filled with hypercharge bosons. These hypercharge bosons can very efficiently scatter into Higgs bosons, which can in turn produce the entirety of the particle content in the Standard Model with similarly high efficiency. Tachyonic preheating of hypercharge bosons can thus lead to the inflaton reheating into a plasma of charged particles almost instantaneously.  This results in a very high reheat temperature, which can effectively boost the value of the magnetic field measured today to $B_{\rm eff} \sim 10^{-16}-10^{-15}$ G. The exact value of the axion-gauge coupling does not significantly affect the resulting magnetic field, provided it is large enough to put the system in the regime of instantaneous preheating. This is intuitively understood, since in the case of instantaneous preheating the entirety of the energy-density of the inflaton is transferred to the gauge field modes after the end of inflation. Further increasing the coupling cannot increase the magnetic field since its amplitude is effectively saturated by the total available energy density of the inflaton. However, if the coupling increases beyond $\alpha/f \sim 65 m_{\rm pl}^{-1}$, the backreaction becomes large enough to trap the axion \emph{during inflation}, which momentarily stops rolling down its potential. Inflation stops and, after a brief pause, restarts. This leads to a prolonged period of inflation with the possibility of enhanced primordial black hole production. We leave the study of this effect for future work.

The two main issues regarding cosmological magnetic fields is their presence in galaxies and intergalactic voids. Our model can produce $B_{\rm eff} \sim 10^{-16}-10^{-15}$ G for large couplings. Under normal assumptions about the evolution of the Universe, this can be translated into magnetic fields with an amplitude of about $10^{-13}$ G and a correlation length of about $10$ pc. Their correlation length is thus below the typical galactic scale, which makes them unlikely candidates for the seeds to the galactic dynamo, which would result in the galactic magnetic fields measured today. However, they can be relevant in the case of intergalactic magnetic fields (IGMF's). Direct observations of distant blazars \cite{Taylor:2011bn} provide a lower bound on IGMF's, depending on their correlation length. 
For magnetic fields with a correlation length less than $1$ Mpc, blazar observations provide limits on a combination of the physical magnetic field $B$ and the corresponding correlation length $\lambda$ through $B_{\rm eff} = B \sqrt {\lambda / 1{\rm Mpc}} $.
However for correlation lengths larger than $1$ Mpc the limit on the field strength is smaller, namely $B  = B_{\rm eff}$.  In both cases $B_{\rm eff} \gtrsim 10^{-17}\, {\rm G}$ or $B_{\rm eff} \gtrsim 10^{-15}\, {\rm G}$ depending on assumptions.
At this point it is important to note the use of the diffuse gamma ray signal instead of direct blazar observations that allows direct probing of the magnetic field spectrum, in the case of helical magnetic fields.
 Recent analyses \cite{Tashiro:2013ita,Chen:2014qva} have inferred the amplitude of the magnetic field at $10$ Mpc to be $B \sim 5.5\times 10^{-14} \, {\rm G}$. Careful analysis of the evolution of the full magnetic field spectrum through the cosmic history, including the relevant MHD simulations, are needed to compare the model predictions with the observed late-time magnetic field strength at $10$ Mpc.

Despite the small correlation length, both the large amplification of the hypermagnetic fields, as well as the very fast transition of the Universe from inflation to a charged plasma allowing for an inverse cascade process, lead to considerably larger $B_{\rm eff}$ than has been estimated before for this type of models. Depending on one's assumption on the mechanism behind the suppression of the cascade signal from distant blazars, the magnetic field produced by this simple model is close or within observed bounds for the IGMF's in the large coupling regime. 
Furthermore, the physical intensity of the produced magnetic field is high enough for large couplings to trigger baryogenesis through the chiral anomaly of the Standard Model \cite{Fujita:2016igl}.

Although we only used the quadratic potential form for the inflaton, which is now in significant tension with cosmic microwave background data, we do not expect our results for the magnetic field to vary significantly for different potentials, provided the energy scale of inflation is comparable. As shown in ref.\  \cite{Adshead:2015pva},  the preheating efficiencies for quadratic and axion-monodromy inflation are qualitatively similar. The quantitative difference lies in the value of the axion-gauge coupling $\alpha/f$ that leads to complete preheating. This can be easily understood, since the tachyonic amplification of the gauge fields is controlled by the parameter $\xi = 0.5 (\alpha / f) (\dot \phi /H)$. Less steep inflationary potentials lead to a smaller velocity $\dot \phi /H$, which must be compensated by increasing the coupling $\alpha/f$ to get comparable gauge field production. However,  instantaneous preheating makes the estimation of the resulting magnetic field both straightforward and robust to changes in the potential. The energy density (equivalently the amplitude) of the hypermagnetic field is largely set by the energy density in the inflaton condensate, in other words the Hubble scale at the end of inflation. Our results can thus be easily transferred to similar high-scale axion inflation models, like axion monodromy.

A further qualitative difference between the quadratic cases considered here and axion monodromy is the production of oscillons at the early stages of preheating. This is due to the fact that the axion-monodromy potential is less steep than a quadratic potential at field values $\phi \sim m_{\rm pl}$ \cite{Amin:2011hj}.  The interplay of multiple fields, scalar and gauge fields, in the context of oscillons, has been studied for a Higgs $SU(2)$ system \cite{Graham:2007ds,sfakianakis:2012bq}. If oscillons are to be considered a realistic possibility for the early Universe, the study of more general models is needed, especially due to the possible complexity of interacting fields in the preheating era, where oscillons are believed to arise.

%The connection of axion inflation to baryogenesis, either through the gauge field \cite{Vachaspati:2001nb} or direct coupling to fermions \cite{Adshead:2015kza, Adshead:2015jza},  could be considered parallel to magnetogenesis but this is beyond the scope of the present paper. 
%%%%%%%%%%%%%%%%%%%%%%%%%%%%%%%%%%%%%%%
%%%%%%%%%%%%%%%%%%%%%%%%%%%%%%%%%%%%%%%
%%%%%%%%%%%%%%%%%%%%%%%%%%%%%%%%%%%%%%%

\acknowledgments
We thank Andrew Long for comments on an early draft and discussions on the relation between primordial magnetic fields and baryogenesis. We thank Jessie Shelton, Charles Gammie, and Daniel Chung for helpful discussions. PA is supported by the United States Department of Energy, DE-SC0015655. JTG is supported by the National Science Foundation, PHY-1414479.  We acknowledge the  National Science Foundation, the Research Corporation for Science Advancement and the Kenyon College Department of Physics for providing the hardware used to carry out these simulations. EIS gratefully acknowledges support from a Fortner Fellowship at the University of Illinois at Urbana-Champaign.

\appendix

\section{Gauge fields during and after axion inflation}
\label{app:gaugefieldsaxion}

In this appendix, we gather some known results about gauge fields during and after axion inflation.

\subsection*{Background and equations of motion}

The equation of motion for the pseudo-scalar field is the Klein-Gordon equation sourced by the Chern-Simons density of the gauge field
\begin{align}
\label{axioneom}
(\partial_\tau^2 + 2\mathcal{H}\partial_\tau - \partial_i \partial_i)\phi + a^2 \frac{dV}{d\phi} = \frac{\alpha}{4 f}a^2 F_{\mu\nu}\tilde{F}^{\mu\nu},
\end{align}
where, $\tau$ is conformal time and  $\mathcal{H} = a'/a$. Here and throughout this appendix a prime represents a derivative with respect to conformal time, $' \equiv \partial_\tau = \partial/\partial \tau$.  We defined $\tau$ to be a negative, increasing quantity during inflation
\begin{align}
d\tau = \frac{dt}{a}, \quad \tau  = & \int_t \frac{dt}{a} = \int \frac{d\ln a}{a H}\approx - \frac{1}{aH},
\end{align}
where the last approximation is exact in the de-Sitter limit, $\epsilon_H \to 0$, where the slow-roll parameter, $\epsilon_H$, is defined as $\epsilon_H = -\dot{H}/H^2$.  Again, in this appendix, an overdot is used to denote a derivative with respect to cosmic time, $t$. 

The equations of motion for the gauge field are 
\begin{align}
\partial_{\rho}\(\sqrt{-g}F^{\rho\sigma}\)+\frac{\alpha}{f} \partial_{\rho}(\sqrt{-g}\phi \tilde F^{\rho\sigma}) = 0.
\end{align}
The $\sigma = 0$ equation is the Gauss' law constraint
\begin{align}
\label{gaugeeom1}
\partial_j \partial_j A_0  - \partial_\tau  \partial_{i} A_i+\frac{\alpha}{f}\epsilon_{ijk} \partial_{k}\phi \partial_i A_j  = 0,
\end{align}
while the $\sigma = i$ equations are the field equations for the spatial components of the gauge field
\begin{align}%\nn
\label{gaugeeom2}
-\partial_{\tau}\(  \partial_\tau A_i - \partial_i A_0 \)+\partial_{m} (\partial_m A_i - \partial_i A_m)+\frac{\alpha}{f} \epsilon_{imk}\partial_{\tau}\phi \partial_m A_k  %& \\ 
-\frac{\alpha}{f} \epsilon_{ i m k}\partial_{m}\phi (\partial_\tau A_k - \partial_k A_0) =  & 0.
\end{align}
Finally, assuming the metric is unperturbed, the scale factor satisfies Einstein's equations,
\begin{equation}
\label{ffriedman}
\frac{3 m_{\rm pl}^{2} }{8\pi}\mathcal{H}^2 =  a^2\rho,
\quad 
\frac{m_{\rm pl}^{2}}{8\pi}\(\mathcal{H}' - \mathcal{H}^2\) =  -a^2\frac{\rho+p}{2}.
\end{equation}
The pressure, $p$, and energy  density, $\rho$, are found from the stress-energy tensor
\begin{align}
 T_{\mu\nu} = &\tr\[F_{\mu\alpha}F_{\nu\beta}\]g^{\alpha\beta}  -\frac{g_{\mu\nu}}{4}F_{\mu\nu}F^{\mu\nu}  -g_{\mu\nu}\left[\frac{1}{2}g^{\rho\sigma}\partial_{\rho}\phi\partial_{\sigma}\phi+V(\phi)\right] + \partial_{\mu}\phi\partial_{\nu}\phi,
\end{align}
which can be explicitly written as
\begin{align}
\rho = & \frac{1}{2}\frac{\phi'{}^2}{a^2} +\frac{1}{2}\frac{(\partial_i \phi)^2 }{a^2}+  V(\phi) + \frac{1}{2 a^4}(\partial_0 A_i - \partial_i A_0)^2 + \frac{1}{4 a^4}(\partial_i A_j - \partial_j A_i)^2
\end{align}
and
\begin{align}
p  = &  \frac{1}{2}\frac{\phi'{}^2}{a^2} +\frac{1}{2}\frac{(\partial_i \phi)^2 }{a^2}-  V(\phi) + \frac{1}{6 a^4}(\partial_0 A_i - \partial_i A_0)^2 + \frac{1}{12 a^4}(\partial_i A_j - \partial_j A_i)^2.
\end{align}
Note that the axion-gauge field coupling does not contribute directly to the stress-energy tensor.

\subsubsection*{Gauge-field production during inflation}

We work in the Coulomb (or transverse) gauge $\partial_i A_i=0$.  At linear order in fluctuations, this gauge choice along with the Gauss' law constraint of eq.\ \eqref{gaugeeom1} implies that $A_0=0$.  With the approximation of de-Sitter space and constant $\dot{\phi}/H$, one can  solve the equations of motion for the gauge fields during inflation.

At linear order in fluctuations, in Coulomb gauge, the equation of motion for the gauge field becomes
\begin{align}\label{eqn:gauge field}
\partial^2_{\tau}A_i - \partial_m\partial_m A_i - \frac{\alpha}{f}\epsilon_{imk}\partial_\tau \phi \partial_m A_k = 0 .
\end{align}
We work in Fourier space to quantize the gauge field, and expand each Fourier mode in helicity states %
\begin{align}
\vec{A}({\bf x})  = & \int \frac{d^3 k}{(2\pi)^3} \vec{A}_{\bf k}  e^{i {\bf k}\cdot \bf{x}},\quad 
%\end{align}
%
%\begin{align}
\vec{A}_{\bf k} = \sum_{\lambda = \pm}A^{\lambda}_{\bf k} \vec{\varepsilon}
\,{}^{\lambda}({\bf k }),
\end{align}
where the polarization vectors, $ \varepsilon^{\lambda}_i({\bf k })$, denoting transverse left- and right-handed polarized waves, satisfy the orthogonality and normalization relations
\begin{align}\nn\label{eqn:polvecs}
k_i \varepsilon^{\pm}_{i}({\bf k}) = & 0,\quad
\epsilon^{ijk}k_{j} \varepsilon^{\pm}_{k}({\bf k})  =  \mp i k \varepsilon^{\pm}_{i}({\bf k}), \\
\varepsilon_i^\pm({\bf k})^* = & \varepsilon_i^{\pm}(-{\bf k}), \quad 
 \varepsilon^{\lambda}_{i}({\bf k})\varepsilon^{\lambda'}_{i}(-{\bf k})  =  \delta_{\lambda \lambda'}.
\end{align}
The longitudinal modes of the gauge field not accounted for by this helicity decomposition. This is consistent with a linear-order analysis, since the longitudinal part scales like the product of $\nabla \phi$ and the gauge field (see eq.\ \eqref{gaugeeom1}) and  is therefore higher-order in fluctuations. 

The modes are quantized by introducing the creation and annihilation operators, $a_{\lambda}({\bf k})$ and $a_{\lambda}^{\dagger}({\bf k})$,  satisfying the canonical commutation relations
\begin{align}
\[a_{\lambda}({\bf k}), a^{\dagger}_{\lambda'}({\bf k'})\] = (2\pi)^{3}\delta_{\lambda\lambda'}\delta^{3}({\bf k} - {\bf k}'),
\end{align}
which allows us to expand the mode-functions as
\begin{align}
A_i(\tau, {\bf x}) = & \sum_{\lambda = \pm}\int \frac{d^3 k}{(2\pi)^3}e^{i {\bf k}\cdot {\bf x}}\varepsilon^{\lambda}_{i}({\bf k}) \[A^{\lambda}(k, \tau) a_{\lambda}({\bf k}) +A^{\lambda,*}(k, \tau)a^{\dagger}_{\lambda}(-{\bf k}) \].
\end{align}
With our conventions, the gauge field equation of motion eq.\ \eqref{gaugeeom2} becomes a separate equation for each polarization, depending only on the magnitude of the momenta $k = |{\bf k}|$,
\begin{align}\label{eqn:kspacegfeqn}
\(\partial_{\tau}^2 + k^2 \pm 2\xi (aH\tau)\frac{k}{\tau}\) A^{\pm}_{k}= 0, \quad \gfc =  \frac{1}{2}\frac{\alpha}{f}\frac{\dot{\phi}}{H} = {\rm sign}(\dot\phi) {m_{\rm pl}\over \sqrt{8\pi}} \frac{\alpha}{f}\sqrt{\frac{\epsilon_H}{2}}.
\end{align}
The effective coupling strength, $\xi$, controls gauge field production during inflation. In the de-Sitter limit, where $(aH\tau) = -1$ and $\xi$ is constant, the exact solution can be written in terms of the Whittaker W-function. Compared to the conformally invariant radiation solution, the relative amplification of each circularly polarized mode is given by 
\begin{align}
\left| \frac{A^{\pm}}{A^{\pm,\rm rad}}\right| \simeq e^{\frac{\pi}{2}|\gfc|\pm \frac{\pi}{2}\gfc} \, ,\quad |\gfc| > 1.
\label{eq:relativeamplification}
\end{align}
Note that, for $\gfc > 0$ ($\gfc < 0$), the mode $A^{+}_k$ ($A^{-}_k$) gets amplified by a factor $\sim e^{\pi |\gfc|}$ while the other helicity mode is unchanged. Without loss of generality, we assume that $\dot\phi < 0 $, which leads to the negative helicity modes being amplified during inflation. As in \cite{Adshead:2015pva} we focus on large-field inflationary models, where the axion shift-symmetry is theoretically well motivated.

\subsubsection*{Gauge field backreaction during inflation}

The exponentially enhanced gauge fields have important effects during inflation due to their re-scattering off the inflaton condensate and their interactions with the metric. As inflation progresses, the field velocity measured in units of the Hubble rate, $|\dot\phi|/H\propto \sqrt{\epsilon_H}$, increases. This means that shorter-wavelength modes that leave the horizon later during inflation are amplified more than their longer-wavelength counterparts that leave the horizon earlier. The largest effects occur when $\epsilon_H$ is near unity near the end of inflation.  The former leads to the production of fluctuations of the inflaton which are statistically non-Gaussian, while the latter leads to the production of gravitational radiation \cite{Barnaby:2011qe}. Based on the bounds on non-Gaussianities in the CMB from Planck  \cite{Ade:2013ydc}, the axion-gauge coupling can be constrained to be $\gfc_{\rm CMB} \lesssim 2.22$, where $\gfc_{\rm CMB}$ is the quantity defined in eq.\ \eqref{eqn:kspacegfeqn} evaluated during the time when the CMB-relevant modes leave the horizon.

In the limit that $\gfc \gg 1$ (which for models that satisfy $\gfc_{\rm CMB} < 2.22$ only possibly occurs near the end of inflation), the energy density in the gauge fields becomes important and the gauge-field fluctuations begin to backreact on the homogeneous background equations of motion. In this limit, using the Hartree approximation, the Friedmann (eq.~\eqref{ffriedman}) and Klein-Gordon equations (eq.\ \eqref{axioneom}) become
\begin{equation}
\label{eqn:backreactfried}
\frac{3 m_{\rm pl}^{2} }{8\pi}\mathcal{H}^2 =  \frac{\phi'{}^2}{2}+a^2V(\phi) + \frac{a^2}{2}\langle E^2+B^2\rangle,
\end{equation}
\begin{equation}
\frac{m_{\rm pl}^{2}}{8\pi}\(\mathcal{H}' - \mathcal{H}^2\) =   -\(\frac{\phi'{}^2}{2} +\frac{2}{3}a^2\langle E^2+B^2\rangle\),
\label{eqn:friednmaneq}\end{equation}
\begin{equation}
\label{eqn:backreactKG}
{\phi}''+2\mathcal{H}\phi' + a^2V' =  \frac{\alpha}{f} a^2 \langle {\bf E} \cdot {\bf B} \rangle,
\end{equation}
where the electric and magnetic fields associated with the $U(1)$ gauge field are $E_i = a^{-2}A'_i$ and $B_i = a^{-2} \epsilon_{ijk}\partial_j A_k$. In this limit, up to an irrelevant constant phase, the gauge field mode that is amplified is approximated near horizon crossing by \cite{Anber:2009ua}
\begin{align}
A^{-}_k (\tau) = \frac{1}{\sqrt{2k}}\(\frac{k|\tau|}{2|\gfc|}\)^{1/4}\exp\(\pi|\gfc| - 2\sqrt{2|\gfc| k|\tau|} \),
\label{eq:approximateAk}
\end{align}
while the other mode is unaffected and is negligible. The expectation values of the quantum fields are well approximated by \cite{Anber:2009ua}.
\begin{align}\label{eqn:GFenergy}
\frac{1}{2}\langle E^2+B^2\rangle \simeq & 1.4 \cdot 10^{-4}\frac{H^4}{|\gfc|^3}e^{2\pi| \gfc|}, \quad
\langle {\bf E} \cdot {\bf B} \rangle \simeq  2.4 \cdot 10^{-4} \frac{H^4}{|\gfc|^4}e^{2\pi|\gfc|}.
\end{align}
Toward the end of inflation, for large values of $m_{\rm pl}\, \alpha/f$, the backreaction of the gauge fields on the rolling axion becomes important and inflation is prolonged \cite{Barnaby:2011qe}. During this phase, the primordial density fluctuation spectrum is expected to be dominated by rescattering and large, non-Gaussian density fluctuations are predicted.

The backreaction of the produced gauge fields on the inflaton spectrum during inflation can be approximately calculated from 
\begin{align}
\left [ \partial_\tau^2 +k^2 -{a''\over a} + a^2 m^2 \right ] (a\delta\phi) = {\alpha \over f} a^3 \left ( E\cdot B -  \langle E\cdot B \rangle \right ).
\label{eq:deltaphieom}
\end{align}
Following the discussion found for example in \cite{Barnaby:2011vw, Fujita:2015iga}, we can formally solve eq.\ \eqref{eq:deltaphieom} as
\begin{align}
a \delta\phi(k,\tau) =Q_k(\tau) +  2\int_{-\infty}^{\tau} d\tau' \, \Im[Q^*_k(\tau') Q_k(\tau) ] J_{\rm EM} (k,\tau') ,
\label{eq:deltaphi_QJ}
\end{align}
where $Q_k(\tau)$ is the homogenous solution of eq.\ \eqref{eq:deltaphieom}. 

By using the de-Sitter approximation for the conformal time $\tau = -1/aH$ and neglecting the mass term since $m \ll H$, we can re-write eq.\ \eqref{eq:deltaphieom} as
\begin{align}
\left [ \partial_\tau^2 +k^2 -{2\over \tau^2}  \right ] (a\delta\phi) = {\alpha \over f} a^3 \left ( E\cdot B -  \langle E\cdot B \rangle \right )
\label{eq:deltaphieom}
\end{align}
where the homogenous solution is simply
\begin{align}
Q_k(\tau)  = {1\over \sqrt{2k}} \left ( 1 - {i\over k\tau} \right ) e^{-ik\tau}.
\end{align}
We can now calculate the inflaton perturbations $\delta \phi$ from eq.\ \eqref{eq:deltaphi_QJ}, where
\begin{align}
J_{\rm EM} = a^3 {\alpha\over f} \int {d^3 p \over (2\pi)^3}\,  \vec E\big ( \vec p,\tau \big ) \cdot \vec B\big ( \vec k-\vec p,\tau \big ).
\end{align}
The power spectrum of $\delta \phi$, as given for example in \cite{Barnaby:2011vw, Fujita:2015iga} is
\begin{align}
\nonumber
{\cal P}_{\delta \phi}(k,\tau) &= {\alpha^2 \over f^2} {k^3 \over 2\pi^2 a^2 }  \int {d^3p\over (2\pi)^3}  \left ( 1-  \hat p \cdot \widehat{k-p} \right ) 
\\
\times &\left [ p^2 | {\cal I} (\tau, k ; p , |k-p|^2 ) |^2 +  p |k-p|  {\cal I} (\tau, k ; p , |k-p|^2 ) {\cal I}^* (\tau, k ; p , |k-p|^2 ) \right ],
\label{eq:deltaphi_spectrum}
\end{align}
where
\begin{align}
{\cal I} (\tau, k ; p,q) = \int_{-\infty}^{\tau} {d\tau' \over a(\tau')} \Im[Q_k(\tau') Q^*_k(\tau)] \, A_p(\tau') A'_q(\tau').
\end{align}
and $A_k(\tau)$ is the gauge field mode $A^-_k$.
Comparing this expression to the one given in ref.\ \cite{Fujita:2015iga}, we see that we are missing the summation over the gauge field polarization.  We are justified in only considering the backreaction of this polarization onto the inflaton because only one polarization is amplified during inflation, as in ref.\ \cite{Barnaby:2011vw}.  The relevant amount of backreaction is estimated in section \ref{sec:LatticeSimulations}.

\section{Sampling effects}
\label{app:sampling}

In field theory we can formally decompose a real scalar field using a continuum of creation and annihilation operators as
\beq
\delta\phi(x,t) = \int {d^3 k \over (2\pi)^3} \left  [ a_k e^{i\omega t} e^{-i \vec k \cdot \vec x} + a_k^\dagger e^{-i\omega t} e^{i \vec k \cdot \vec x} \right ]
\eeq
where we took the background to correspond to Minkowski space-time. However, when initializing and evolving a scalar field on a grid, we cannot use creation and annihilation operators, but instead we use Gaussian random variables---their classical counterparts.  The mode functions, $f_k$, can be decomposed onto the forward-moving and backward-moving parts,
\beq
f_k(t) = f_k^R e^{-ikt} + f_k^L e^{ikt}.
\label{modedecomposed}
\eeq
If we use eq.\ \eqref{modedecomposed} to calculate the classical power in this mode we get
\beq
|f_k|^2 _{\rm class}   =  |f_k^R|^2 +|f_k^L|^2   + f_k^L \bar f_k^R e^{2ikt} + f_k^R \bar f_k^L e^{-2ikt},
\eeq
where $f_k^{R,L}$ are the classical analogues of $a_k$ and $a_k^\dagger$. If we want to move from a quantum operator to a classical observable, we must define the vacuum expectation value of the operator, as
\beq
|f_k|^2 _{\rm class}  = \langle 0 | |f_k|^2_{\rm quant} | 0\rangle
\eeq
In a field theory calculation, the terms that are proportional to $e^{\pm 2ik t}$ simply vanish, due to the creation and annihilation operators annihilating the left or right vacuum states. 
In the case of a lattice calculation, computing the vacuum expectation value corresponds to averaging over different realizations of a particular wavenumber $|\vec k|$. For any finite number of modes, this calculation includes an uncertainty, in the sense that the terms proportional to $e^{\pm 2ik t}$ would average to zero if we had access to an infinite number of independent modes, but will otherwise give a finite oscillatory contribution.

In order to present the results of our simulation, we collect all independent wavenumbers and bin them in intervals of $\Delta k =2\pi/L \approx 0.4\,m$. The number $N$ of independent modes in each bin $k$ scales as $ k \sim N^2$ for low $k$ (until the magnitude of the wavenumber reaches $k = 120\times 2\pi/L \approx 50\,m$ at which point the number of modes in each bin begins to decrease) which is the known result for the density of states in a spherical shell in three dimensions. We have checked that the relative amplitude of the early-time oscillations in the power of different modes shown in figure \ref{fig:BD_deltaphi} scales as $1/\sqrt N$, as expected. Varying the starting time introduces a phase-shift into these early-time oscillations, which is responsible for the small differences in the final results shown in figure \ref{fig:varyNstart}. This is an inherent and well understood statistical effect in the simulation, but one in which the reader might have interest.

\bibliographystyle{JHEP}
\bibliography{AxMagnet}

\end{document}